\documentclass[11pt]{article}

\usepackage{amsfonts}
\usepackage{makeidx}
\usepackage[totalwidth=460truept,totalheight=600truept]{geometry}
\usepackage{latexsym,amssymb,amsmath,graphicx,accents,eucal,slashed,subfigure}
\usepackage{dsfont}
\usepackage[T1]{fontenc}
\usepackage{hyperref}

\global\arraycolsep=1truept

\numberwithin{equation}{section}

\begin{document}

\bigskip \phantom{C}

\vskip 1.4truecm

\begin{center}
{\huge \textbf{Weighted Power Counting And}}

\vskip .5truecm

{\huge \textbf{Chiral Dimensional Regularization}}

\vskip 1truecm

\textsl{Damiano Anselmi}

\vskip .2truecm

\textit{Dipartimento di Fisica ``Enrico Fermi'', Universit\`{a} di Pisa, }

\textit{and INFN, Sezione di Pisa,}

\textit{Largo B. Pontecorvo 3, I-56127 Pisa, Italy}

\vskip .2truecm

damiano.anselmi@df.unipi.it

\vskip 1.5truecm

\textbf{Abstract}
\end{center}

\medskip

We define a modified dimensional-regularization technique that overcomes
several difficulties of the ordinary technique, and is specially designed to
work efficiently in chiral and parity violating quantum field theories, in
arbitrary dimensions greater than 2. When the dimension of spacetime is
continued to complex values, spinors, vectors and tensors keep the
components they have in the physical dimension, therefore the $\gamma $
matrices are the standard ones. Propagators are regularized with the help of
evanescent higher-derivative kinetic terms, which are of the Majorana type
in the case of chiral fermions. If the new terms are organized in a clever
way, weighted power counting provides an efficient control on the
renormalization of the theory, and allows us to show that the resulting 
\textit{chiral dimensional} \textit{regularization} is consistent to all
orders. The new technique considerably simplifies the proofs of properties
that hold to all orders, and makes them suitable to be generalized to wider
classes of models. Typical examples are the renormalizability of chiral
gauge theories and the Adler-Bardeen theorem. The difficulty of explicit
computations, on the other hand, may increase.

\vskip 1truecm

\vfill\eject

\section{Introduction}

\label{RGren}\setcounter{equation}{0}

The dimensional regularization \cite{dimreg,dimreg2} is very convenient to
make calculations in quantum field theory and prove properties to all
orders, such as renormalizability, when gauge anomalies are manifestly
absent. It has some annoying features in chiral, parity violating and
sypersymmetric theories, where the matrix $\gamma _{5}$ and the tensor $%
\varepsilon _{\mu \nu \rho \sigma }$ play important roles. There, manifest
gauge invariance is lost, and calculations are involved. More importantly,
the proofs of all-order properties are quite demanding, because several
common arguments do not work and some classifications of counterterms are
ambiguous.

According to the standard definition of dimensional regularization for
chiral theories in four dimensions \cite{dimreg,maison}, calling $%
D=4-\varepsilon $ the continued, complex dimension of spacetime, the $D$%
-dimensional spacetime manifold $\mathbb{R}^{D}$ is broken into the product $%
\mathbb{R}^{4}\times \mathbb{R}^{-\varepsilon }$ of the physical spacetime
manifold $\mathbb{R}^{4}$ times an ``evanescent'' space, $\mathbb{R}%
^{-\varepsilon }$. The $D$-dimensional $\gamma $ matrices are formal objects
that satisfy the continued Dirac algebra $\{\gamma ^{\mu },\gamma ^{\nu
}\}=2\eta ^{\mu \nu }=2\hspace{0.01in} $diag$(1,-1,\cdots ,-1)$. Then, the
matrix $\gamma _{5}$ is formally defined as the product $i\gamma ^{0}\gamma
^{1}\gamma ^{2}\gamma ^{3}$ \cite{dimreg}. This approach, due to 't Hooft
and Veltman, is consistent to all orders, gives the right one-loop anomalies
and can be easily generalized to arbitrary even dimensions. Although it is
widely used, it has some undesirable features. The most important one, in
our viewpoint, is that it is responsible for various obstacles that obstruct
the proofs of theorems to all orders in perturbation theory. Many
alternative approaches have been proposed in the literature (see ref. \cite%
{alternatives} for a nonexhaustive list of recent contributions to this
subject), mainly focused on the definition of $\gamma _{5}$. However, most
of those proposals do not simplify the proofs of all-order theorems.

The true origin of the difficulties we are concerned with is not $\gamma
_{5} $ \textit{per se} but the continued Dirac algebra. In this paper we
show that it is possible to avoid most inconveniences by working with the
usual $d$-dimensional $\gamma $ matrices in arbitrary $D$ dimensions, where $%
d$ denotes the dimension of physical spacetime. To achieve this and other
goals, it is necessary to upgrade the whole dimensional regularization to a
new technique, which we call \textit{chiral dimensional (CD) regularization}.

The first problem we must face is the regularization of chiral fermions. The
na\"{\i}ve $D$-dimensional continuation of their action is gauge invariant,
but does not provide good propagators. In the CD regularization this problem
is solved by adding evanescent kinetic terms of the Majorana type. Since no
evanescent $\gamma $ matrices are allowed, the evanescent kinetic terms must
be \textit{higher derivative}, which means that they are multiplied by
parameters of negative dimensions in units of mass. In general, those
parameters may propagate into the physical sector of the theory, turn
nonrenormalizable vertices on and cause all other sorts of troubles. To keep
the evanescent sector under control and prove that the CD regularization is
consistent to all orders, we arrange the regularization technique so that it
satisfies the requirements of weighted power counting \cite{halat}. Doing
so, we obtain an effective control over the locality of counterterms and the
renormalization to all orders.

The propagators of all fields must be corrected similarly, to make their
structure match the structure of fermionic propagators. Since gauge
invariance is not (and does not need to be) preserved away from $d$
dimensions, we show that it is consistent to require that all fields have
strictly $d$-dimensional components. We use this property to simplify the
regularization technique as much as we can.

Summarizing, in the CD regularization all fields (scalars, spinors, vectors,
tensors) have exactly the same components in $D$ dimensions, as they have in 
$d$ dimensions, and the $\gamma $ matrices, as well as the $\varepsilon
_{\mu _{1}\cdots \mu _{d}}$ tensors, are just the usual $d$-dimensional
ones. Evanescent terms are added, guided by weighted power counting, to make
propagators fall off with appropriate velocities in all directions of
integration.

The CD\ regularization is particularly convenient to treat general gauge
theories that cannot be regularized preserving their gauge symmetries in a
manifest way. Examples are chiral theories and parity violating theories,
such as the standard model, Lorentz violating extensions of the standard
model \cite{colladay,lvsm,noh}, Chern-Simons theories in three dimensions,
and so on. In the realm of nonrenormalizable theories we mention the
standard model coupled with quantum gravity, as well as supergravity.

The CD regularization keeps the good properties of the dimensional
regularization, among which the fact that local perturbative changes of
field variables have Jacobian determinants identically equal to one, which
simplifies the Batalin-Vilkovisky master equation \cite{bata} and several
derivations. At the same time, it overcomes known and less-known
inconveniences of the usual dimensional regularization, such as problems due
to the dimensional continuation of Fierz identities and the propagation of
evanescences through the Batalin-Vilkovisky antiparentheses \cite{bata}.
Ultimately, the new technique provides a powerful tool to make systematic
proofs to all orders with a relatively small effort in perturbative quantum
field theory, in arbitrary dimensions $d>2$.

One situation where the problems of the common dimensional regularization
can be appreciated is the proof of the Adler-Bardeen theorem \cite%
{adlerbardeen,review}, which is a realm where the most advanced techniques
of perturbative quantum field theory must be used altogether. The original
proof given by Adler and Bardeen \cite{adlerbardeen} works only in QED. Most
generalizations to non-Abelian gauge theories use\ arguments based on the
renormalization group \cite{zee,collins,tonin,sorella}. However, those
arguments have some limitations, because they do not apply to conformal
field theories, finite theories or theories where the first coefficients of
the beta functions vanish \cite{sorella}. Algebraic/geometric derivations 
\cite{witten} based on the Wess-Zumino consistency conditions \cite%
{wesszumino} and the quantization of the Wess-Zumino-Witten action also do
not seem suitable to be generalized. Recently, a more powerful proof of the
cancellation of gauge anomalies to all orders, when they vanish at one loop,
was given in ref. \cite{ABrenoYMLR}, elaborating on arguments that first
appeared in ref. \cite{lvsm}. That approach has the virtue of identifying
the subtraction scheme where the cancellation to all orders is manifest.
Nevertheless, due to the difficulties of the dimensional regularization, it
only covers particular classes of models, namely four dimensional,
perturbatively unitary power counting renormalizable gauge theories coupled
to matter, where it is possible to handle the inconveniences in \textit{ad
hoc} ways. It would be important to extend those results to all perturbative
quantum field theories, renormalizable or not. Going through the arguments
of ref. \cite{ABrenoYMLR}, it is easy to spot several steps that do not
generalize to wider classes of models in a straightforward way, and
understand that the main obstacle is indeed the dimensional regularization
as we normally understand it. On the other hand, when it comes to proving
properties to all orders in perturbation theory, no known regularization
technique is more powerful than the dimensional one, for a variety of
reasons. Thus, to move forward it is necessary to formulate a more versatile
regularization technique that overcomes the difficulties, but keeps the
benefits of the ordinary dimensional one. The CD regularization provides a
satisfactory answer to this problem.

It is known that in parity violating theories algebraic manipulations and
evaluations of Feynman diagrams are more demanding than in parity invariant
theories. Using the CD regularization the difficulty of explicit
computations may increase. Nevertheless, we do not worry about this problem
here, because we think that a certain complexity is unavoidable and a
reasonable price to pay, if we want to simplify the proofs of all-order
theorems. If we are interested in simplifying calculations, instead, we must
deform the dimensional regularization in a different way \cite%
{regula3d,regulasm}, by making propagators have $SO(1,D-1)$-scalar
denominators. In an explicit example we show that the effort to compute
one-loop divergent parts and anomalies with the CD regularization is
comparable to the usual one.

The paper is organized as follows. In section 2 we define the chiral
dimensional regularization. We study chiral fermions, scalar fields, gauge
fields, gravity in the metric-tensor formalism, gravity in the vielbein
formalism and Chern-Simons theories. In section 3 we use weighted power
counting to study the locality of counterterms and renormalization, and
prove that the CD regularization is consistent to all orders. In section 4
we calculate the one-loop chiral anomaly with the CD technique. In section 5
we show how to use the new technique in several applications. In particular,
we simplify the classification of counterterms and contributions to
potential anomalies and show how to overcome a number of obstacles that
afflict the proofs of all-order theorems. We concentrate in particular on
the Adler-Bardeen theorem and the proofs of renormalizability of general
gauge theories to all orders, and show that the CD regularization makes the
usual derivations suitable to be generalized to wider classes of models.
Section 6 contains our conclusions.

\section{Chiral dimensional regularization}

\setcounter{equation}{0}

In this section we formulate the new regularization technique, in arbitrary
dimensions $d>2$. As usual, we split the $D$-dimensional spacetime manifold $%
\mathbb{R}^{D}$ into the product $\mathbb{R}^{d}\times \mathbb{R}%
^{-\varepsilon }$ of the ordinary $d$-dimensional spacetime $\mathbb{R}^{d}$
times a residual $(-\varepsilon )$-dimensional evanescent space, $\mathbb{R}%
^{-\varepsilon }$. The spacetime indices $\mu ,\nu ,\ldots $ of vectors and
tensors are split into the bar indices $\bar{\mu},\bar{\nu},\ldots $, which
take the values of $0,1,\cdots ,d-1$, and the formal hat indices $\hat{\mu},%
\hat{\nu},\ldots $, which denote the $\mathbb{R}^{-\varepsilon }$
components. For example, the momenta $p^{\mu }$ are split into the pairs $p^{%
\bar{\mu}}$, $p^{\hat{\mu}}$, also written as $\bar{p}^{\mu }$, $\hat{p}%
^{\mu }$, and the coordinates $x^{\mu }$ are split into $\bar{x}^{\mu }$, $%
\hat{x}^{\mu }$. To evaluate a $D$-dimensional integral we must first
integrate over the hat components of the momenta or coordinates, using the
common formulas of the dimensional regularization, and then integrate over
the bar components. The formal flat-space metric $\eta _{\mu \nu }$ is split
into the usual $d\times d$ flat-space metric $\eta _{\bar{\mu}\bar{\nu}}=$%
diag$(1,-1,\cdots ,-1)$ and the formal evanescent metric $\eta _{\hat{\mu}%
\hat{\nu}}=-\delta _{\hat{\mu}\hat{\nu}}$ (the off-diagonal components $\eta
_{\bar{\mu}\hat{\nu}}$ being equal to zero). When we contract evanescent
components we use the metric $\eta _{\hat{\mu}\hat{\nu}}$, so for example $%
\hat{p}^{2}=p^{\hat{\mu}}\eta _{\hat{\mu}\hat{\nu}}p^{\hat{\nu}}$.

We assume that all of the fields $\Phi (x)$ have strictly $d$-dimensional
components, each of which is a function of $\bar{x}$ and $\hat{x}$. The
spinors $\psi ^{\alpha }$ have $2^{[d/2]}$ components, where $[d/2]$ is the
integral part of $d/2$. Vectors have the $d$ components $A_{\bar{\mu}}$,
with no evanescent component $A_{\hat{\mu}}$ being turned on. Symmetric
tensors have $d(d+1)/2$ components. In particular, the metric tensor $g_{\mu
\nu }$ is made of the diagonal blocks $g_{\bar{\mu}\bar{\nu}}$ and $\eta _{%
\hat{\mu}\hat{\nu}}$, while the off-diagonal components $g_{\bar{\mu}\hat{\nu%
}}$ vanish. Antisymmetric tensors have $d(d-1)/2$ components, and so on.

As anticipated in the introduction, we take the $\gamma $ matrices to be
strictly $d$ dimensional, and satisfy the usual Dirac algebra $\{\gamma ^{%
\bar{\mu}},\gamma ^{\bar{\nu}}\}=2\eta ^{\bar{\mu}\bar{\nu}}$. When we write 
$\gamma ^{\mu }$ we mean $\delta _{\bar{\nu}}^{\mu }\gamma ^{\bar{\nu}}$. If 
$d=2k$ is even, we define the $d$-dimensional generalization of $\gamma _{5}$
as 
\begin{equation*}
\tilde{\gamma}=-i^{k+1}\gamma ^{0}\gamma ^{1}\cdots \gamma ^{2k-1},
\end{equation*}%
which satisfies $\tilde{\gamma}^{\dagger }=\tilde{\gamma}$, $\tilde{\gamma}%
^{2}=1$. Then we have the left and right projectors $P_{L}=(1-\tilde{\gamma}%
)/2$, $P_{R}=(1+\tilde{\gamma})/2$ in the usual fashion. The tensor $%
\varepsilon ^{\bar{\mu}_{1}\cdots \bar{\mu}_{d}}$ and the charge-conjugation
matrix $\mathcal{C}$ also coincide with the usual ones. Full $SO(1,D-1)$
invariance is lost in most expressions, replaced by $SO(1,d-1)\times
SO(-\varepsilon )$ invariance.

These rules are not the end of the story, because once we apply them, we
realize that the so modified dimensional regularization does not equip the
fields with good propagators. For example, the free action 
\begin{equation}
\int \mathrm{d}^{D}x\hspace{0.01in}\ \bar{\psi}_{L}i\bar{\slashed{\partial}}%
\psi _{L}  \label{freea}
\end{equation}%
of the left-handed fermions $\psi _{L}$ gives the propagator 
\begin{equation}
P_{L}\frac{i}{\slashed{\bar{p}}}P_{R},  \label{unpro}
\end{equation}%
which does not fall off in all directions of integration, because it does
not depend on the evanescent components $\hat{p}$ of the momenta. At the
same time, we cannot modify the action (\ref{freea}) in a local way so that
the propagators of chiral fermions get denominators of the form $\bar{p}%
^{2}+\eta \hat{p}^{2}$, where $\eta $ is a positive constant, because the $%
\gamma $ matrices do not have evanescent components, $\gamma ^{\hat{\mu}}$.
These difficulties do not appear with bosonic fields, in general (with the
exception of Chern-Simons gauge fields in three dimensions). However, we
cannot let different fields have propagators with different structures,
because if we did we would not be able to use weighted power counting.

The best we can do is equip all of the fields with propagators that have
denominators equal to products of polynomials 
\begin{equation}
D(\bar{p},\hat{p},m,\varsigma ,\eta )=\bar{p}^{2}-m^{2}-\varsigma \frac{(%
\hat{p}^{2})^{2}}{M^{2}}+\eta \frac{\hat{p}^{2}}{M}+i0,  \label{denni}
\end{equation}
where $\varsigma $ is a positive constant of order one and $M$ is a mass
scale. This property ensures that propagators behave appropriately; in
particular they fall off in all directions of integration, even when $m=\eta
=0$ (which is the case of chiral fermions).

However, formula (\ref{denni}) shows that propagators fall off more rapidly
in the evanescent directions $\hat{p}$ of the momenta than in the physical
directions $\bar{p}$. The right tool to manage different behaviors in
different directions of integration is the \textit{weighted power counting }%
introduced in ref. \cite{halat}. Here we recall its fundamental properties,
leaving other details to section 4. Basically, the structure (\ref{denni})
suggests that in the ultraviolet limit $\bar{p}$ and $\hat{p}^{2}$ should be
regarded as equally important. Ordinary power counting, which is based on
the dimensions in units of mass, instead states that $\bar{p}$ and $\hat{p}$
are equally important in the ultraviolet limit. Thus, we must replace the
dimensions in units of mass with suitable ``weights'', defined in such a way
that $\bar{p}$ and $\hat{p}^{2}$ are equally weighted.

We conventionally take $\bar{p}$ to have weight 1, so the evanescent
components $\hat{p}$ of the momenta have weight 1/2. The action is obviously
weightless, as well as the scale $M$ appearing in formula (\ref{denni}).
Writing the dominant kinetic terms (i.e. the kinetic terms with the largest
number of derivatives $\bar{\partial}$) of the field $\Phi $ as 
\begin{equation}
\frac{1}{2}\int \Phi \bar{\partial}^{N_{\Phi }}\Phi ,\qquad \text{or}\qquad
\int \bar{\Phi}\bar{\partial}^{N_{\Phi }}\Phi ,  \label{dom}
\end{equation}
depending on the case, the weight of $\Phi $ is equal to $(d-N_{\Phi })/2$
and coincides with its dimension in units of mass.

To ensure that propagators are well behaved in all the directions of
integration, we proceed as follows. Consider a polynomial $Q(\bar{p},\hat{p}%
) $ that is also a $SO(1,d-1)\times SO(-\varepsilon )$ scalar. Define its
\textquotedblleft weighted degree\textquotedblright\ as its ordinary degree
once $Q$ is rewritten as a polynomial $\tilde{Q}(\bar{p},\hat{p}^{2})$ of $%
\bar{p}$ and $\hat{p}^{2}$. We require that propagators be rational
functions of the momenta of the form 
\begin{equation}
\frac{P_{2w-N_{\Phi }}^{\prime }(\bar{p},\hat{p})}{P_{2w}(\bar{p},\hat{p})},
\label{propag}
\end{equation}%
where $P_{2w-N_{\Phi }}^{\prime }$ and $P_{2w}$ are $SO(1,d-1)\times
SO(-\varepsilon )$-scalar polynomials of weighted degrees $2w-N_{\Phi }$ and 
$2w$, respectively, such that the monomials $(\bar{p}^{2})^{w}$ and $(\hat{p}%
^{2})^{2w}$ belonging to the denominators $P_{2w}(\bar{p},\hat{p})$ are both
multiplied by nonvanishing coefficients. In the next subsections we show
that it is possible to arrange the regularized action so that all of the
fields fulfill these requirements. The total action is the one that contains
all of the monomials compatible with weighted power counting and the
nonanomalous symmetries of the theory, multiplied by the maximum number of
independent coefficients.

Demanding that the action and the scale $M$ be weightless, we can assign
weights to all of the other parameters. The theories that contain only
parameters of non-negative weights (and are such that propagators fall off
with the correct behaviors in the ultraviolet limit) are renormalizable by
weighted power counting. The theories that contain some parameters of
strictly negative weights are nonrenormalizable. In that case, the
propagators (\ref{propag}) must contain only parameters of non-negative
weights.

Weighted power counting gives us an efficient control on the renormalization
of the CD-regularized theory, including the evanescent sector. It also
ensures that the scale $M$ does not propagate into the physical sector.
Precisely, $M$ is an arbitrary, renormalization-group invariant parameter
that belongs to the evanescent sector from the beginning to the end. In
particular, there is no need to take the limit $M\rightarrow \infty $ at any
stage.

To summarize, to obtain a regularization that effectively works, we must
modify the evanescent sector of the action according to the observations
just made. We begin by showing how this is done in the case of chiral
fermions.

\subsection{Chiral fermions}

The action of the (left-handed) chiral fermions $\psi _{L}$ coupled to gauge
fields in even $d$ dimensions is 
\begin{equation*}
S_{c\psi }=\int \bar{\psi}_{L}i\gamma ^{\bar{\mu}}D_{\bar{\mu}}\psi _{L},
\end{equation*}
where $D_{\bar{\mu}}=\partial _{\bar{\mu}}+gT^{a}A_{\bar{\mu}}^{a}$ is the
covariant derivative and $T^{a}$ are anti-Hermitian matrices associated with
some representation of the gauge group $G$. The propagators (\ref{unpro}) of
this action do not fall off in all directions of integration, because they
are independent of the evanescent components $\hat{p}$ of the momenta. To
overcome this difficulty, we complete the action $S_{c\psi }$ by adding
higher-derivative evanescent kinetic terms of the Majorana type.

For example, in four dimensions, using the standard basis of $\gamma $
matrices (see below) the classical action reads 
\begin{equation}
S_{c\psi }=\int \psi _{L}^{\dagger }i(\partial _{\bar{\mu}}+gT^{a}A_{\bar{\mu%
}}^{a})\bar{\sigma}^{\bar{\mu}}\psi _{L},  \label{scf4}
\end{equation}
where $\bar{\sigma}^{\bar{\mu}}=(1,-\vec{\sigma})$, $\vec{\sigma}$ being the
Pauli matrices. We regularize it by adding the evanescent correction 
\begin{equation}
S_{\text{ev}\psi }=\frac{i}{2M}\int (\varsigma _{\psi }\psi _{L}^{\alpha
}\varepsilon _{\alpha \beta }\hat{\partial}^{2}\psi _{L}^{\beta }+\varsigma
_{\psi }^{*}\psi _{L}^{*\alpha }\varepsilon _{\alpha \beta }\hat{\partial}%
^{2}\psi _{L}^{*\beta }),  \label{sevf}
\end{equation}
where $\varepsilon _{12}=1$. Right-handed chiral fermions can be treated
similarly. Throughout this paper, the constants $\varsigma _{I}$ are
dimensionless and of order one, and multiply the dominant evanescent
quadratic terms. We need a different constant $\varsigma _{I}$ for every
field, because fields renormalize independently of one another, even in the
evanescent sector.

Defining $\Psi =(\psi _{L},\psi _{L}^{*})$, the free propagator of the
action $S_{c\psi }+S_{\text{ev}\psi }$ reads 
\begin{equation}
\langle \Psi (p)\Psi ^{T}(-p)\rangle _{0}=\frac{i}{\bar{p}^{2}-\frac{%
|\varsigma _{\psi }|^{2}}{M^{2}}(\hat{p}^{2})^{2}+i0}\left( 
\begin{tabular}{cc}
$i\epsilon \frac{\varsigma _{\psi }^{*}}{M}\hat{p}^{2}$ & $p_{\bar{\mu}%
}\sigma ^{\bar{\mu}}$ \\ 
$p_{\bar{\mu}}(\sigma ^{\bar{\mu}})^{T}$ & $i\epsilon \frac{\varsigma _{\psi
}}{M}\hat{p}^{2}$%
\end{tabular}
\right) ,  \label{appro}
\end{equation}
where $\epsilon $ is the matrix with entries $\varepsilon _{\alpha \beta }$, 
$\sigma ^{\bar{\mu}}=(1,\vec{\sigma })$ and the superscript $T$ denotes the
transpose.

As promised, the regularized propagators have denominators of the form (\ref%
{denni}), therefore they properly fall off in all directions of integration.
Note that without the regularizing terms (\ref{sevf}), or, equivalently, at $%
\varsigma _{\psi }=0$, the propagators of (\ref{appro}) would be $\hat{p}$%
-independent, and fermionic loops would integrate to zero in dimensional
regularization.

Now we generalize the construction to arbitrary spacetime dimensions $d>2$.
First we need to choose a basis of $\gamma $ matrices. We start from the
usual basis of Pauli matrices $\sigma ^{i}$, $i=1,2,3$ and take $\gamma
^{0}=\sigma ^{1}$, $\gamma ^{1}=i\sigma ^{2}$ in two dimensions. Then we
proceed by induction. Let $\gamma ^{\bar{\mu}}$, $\bar{\mu}=0,1,\ldots 2k-1$%
, denote the $\gamma $ matrices in $d=2k$ dimensions, where $k=1,2,\ldots $.
Define $\gamma ^{\bar{\mu}},\gamma ^{2k}$ as the $\gamma $ matrices in $2k+1$
dimensions, where 
\begin{equation*}
\gamma ^{2k}=-i^{k}\gamma ^{0}\gamma ^{1}\cdots \gamma ^{2k-1},
\end{equation*}
and 
\begin{equation*}
\Gamma ^{0}=1\otimes \sigma ^{1},\qquad \Gamma ^{1}=i\gamma ^{0}\otimes
\sigma ^{2},\qquad \Gamma ^{j+1}=\gamma ^{j}\otimes \sigma ^{2}\text{\quad
for }j=1,\ldots ,2k,
\end{equation*}
as the $\gamma $ matrices in $2k+2$ dimensions.

In this basis, $\gamma ^{\bar{\mu}}$ is always symmetric if $\bar{\mu}$ is
even, and always antisymmetric if $\bar{\mu}$ is odd. The charge-conjugation
matrix 
\begin{equation*}
\mathcal{C}=i^{k(k+1)/2}\gamma ^{0}\gamma ^{2}\cdots \gamma ^{2k-2}
\end{equation*}
in even dimensions $d=2k$ is proportional to the product of the $\gamma $
matrices with even indices, and satisfies 
\begin{equation*}
\mathcal{C}^{\dagger }=-\mathcal{C},\qquad \mathcal{C}^{2}=-1,\qquad 
\mathcal{C}^{T}=(-1)^{k(k-1)/2}\mathcal{C},\qquad \mathcal{C}\gamma ^{\bar{%
\mu}}\mathcal{C}=(-1)^{k}(\gamma ^{\bar{\mu}})^{T},\qquad \mathcal{C}\tilde{%
\gamma}=(-1)^{k}\tilde{\gamma}\mathcal{C}.
\end{equation*}

We generalize the evanescent terms (\ref{sevf}) to even dimensions by
choosing 
\begin{equation}
S_{\text{ev}\psi }=\frac{i}{2M}\int \left( \varsigma _{\psi }\psi _{L}^{T}%
\mathcal{\tilde{C}}\hat{\partial}^{2}\psi _{L}-\varsigma _{\psi }^{*}\bar{%
\psi}_{L}\mathcal{\tilde{C}}\hat{\partial}^{2}\bar{\psi}_{L}^{T}\right) ,
\label{sevfe}
\end{equation}
where $\mathcal{\tilde{C}}$ is a suitable matrix that we now identify. It
must be antisymmetric, otherwise (\ref{sevfe}) vanishes, and invertible, to
give well-behaved propagators. We take a $\mathcal{\tilde{C}}$ that commutes
with $\tilde{\gamma}$, because a $\mathcal{\tilde{C}}$ that anticommutes
with $\tilde{\gamma}$ gives again zero. We also choose $\mathcal{\tilde{C}}$
such that $\gamma ^{0}\mathcal{\tilde{C}}^{\dagger }=\mathcal{\tilde{C}}%
\gamma ^{0}$, so that the expression $S_{\text{ev}\psi }$ given in (\ref%
{sevfe}) is Hermitian. Finally, to simplify a number of manipulations, we
demand that the square of $\mathcal{\tilde{C}}$ be proportional to the
identity. Precisely, we choose $\mathcal{\tilde{C}}^{2}=-1$.

If $d=4$ mod 8 chiral fermions admit Majorana masses, because $[\mathcal{C},%
\tilde{\gamma}]=0$ and $\mathcal{C}$ is antisymmetric. There we can take $%
\mathcal{\tilde{C}}=\mathcal{C}$, which ensures that (\ref{sevfe}) is
Lorentz invariant in the physical portion of spacetime (therefore global
Lorentz symmetry is manifestly nonanomalous). Instead, in even dimensions $%
d\neq 4$ mod 8, chiral fermions cannot have Majorana masses, either because
the charge-conjugation matrix $\mathcal{C}$ anticommutes with $\tilde{\gamma}
$, or because it is symmetric, or both. In even $d>2$, $d\neq 4$ mod 8, we
choose $\mathcal{\tilde{C}}=-i\gamma ^{0}\gamma ^{2}$, which indeed
satisfies $\mathcal{\tilde{C}}^{T}=-\mathcal{\tilde{C}}$, $[\mathcal{\tilde{C%
}},\tilde{\gamma}]=0$, $\mathcal{\tilde{C}}^{\dagger }=-\mathcal{\tilde{C}}$%
, $\{\mathcal{\tilde{C}},\gamma ^{0}\}=0$ and $\mathcal{\tilde{C}}^{2}=-1$.
In this case the regularization explicitly breaks Lorentz symmetry in the
evanescent sector.

In $d=2$ no matrix $\mathcal{\tilde{C}}$ exists, which is why we cannot make
our regularization work there, in general. However, so far we have just
considered single Weyl flavors. With more flavors there are more options
and, depending on the case, it may be possible to build a matrix $\mathcal{%
\tilde{C}}$ that satisfies our requirements even in $d=2$. With more Weyl
flavors Lorentz symmetric evanescent kinetic terms of the Majorana type may
also exist in even dimensions $d\neq 4$ mod 8.

Besides Lorentz symmetry in $d\neq 4$ mod 8, the Majorana terms (\ref{sevfe}%
) may also break other global symmetries, such as those associated with the
gauge groups $G$. However, in most cases such symmetries are nonanomalous,
therefore they can be recovered by adding suitable local counterterms.

Defining $\Psi =(\psi _{L},\bar{\psi}_{L}^{T})$, the propagator of the
action $S_{c\psi }+S_{\text{ev}\psi }$, with $S_{\text{ev}\psi }$ given by (%
\ref{sevfe}), can be worked out with a small effort and reads 
\begin{equation}
\langle \Psi (p)\bar{\Psi}(-p)\rangle _{0}=\frac{i}{D_{p}^{2}+4\tilde{p}%
^{2}|h_{\hat{p}}|^{2}}\left( 
\begin{tabular}{cc}
$D_{p}$ & $2h_{\hat{p}}^{*}\mathcal{\tilde{C}}\tilde{\slashed{p}}$ \\ 
$2h_{\hat{p}}\mathcal{\tilde{C}}\tilde{\slashed{p}}$ & $D_{p}$%
\end{tabular}
\right) \left( 
\begin{tabular}{cc}
$\bar{\slashed{p}}$ & $-h_{\hat{p}}^{*}\mathcal{\tilde{C}}$ \\ 
$-h_{\hat{p}}\mathcal{\tilde{C}}$ & $\bar{\slashed{p}}^{T}$%
\end{tabular}
\right) ,  \label{pevfe}
\end{equation}
where 
\begin{equation*}
D_{p}=\bar{p}^{2}-|h_{\hat{p}}|^{2},\qquad h_{\hat{p}}\equiv i\varsigma
_{\psi }\frac{\hat{p}^{2}}{M},
\end{equation*}
and the vector $\tilde{p}^{\mu}$ is defined as follows. In even dimensions $%
d\neq 4$ mod 8 $\tilde{p}^{\bar{\mu}}$ is obtained from $p^{\bar{\mu}}$ by
suppressing $p^{0}$, $p^{2}$ and $p^{\text{odd}}$, which gives $\tilde{p}^{%
\bar{\mu}}=(0,0,0,0,p^{4},0,p^{6},0,p^{8},\cdots )$. In $d=4$ mod 8,
instead, we take $\tilde{p}^{\bar{\mu}}=0$. Finally, in all cases $\tilde{p}%
^{\hat{\mu}}=0$. It is easy to prove the identities 
\begin{equation*}
\tilde{\slashed{p}}=\tilde{\slashed{p}}^{T},\qquad [\tilde{\slashed{p}},%
\mathcal{\tilde{C}}]=0.
\end{equation*}
Using these relations it is straightforward to verify that (\ref{pevfe}) is
indeed the propagator of the action $S_{c\psi }+S_{\text{ev}\psi }$ for
arbitrary $d>2$.

Formula (\ref{sevfe}) shows that the propagators are well-behaved in all
directions of integrations, according to the requirements formulated in the
previous subsection. Moreover, their denominators are regular everywhere.
This fact can be proved, for example, by checking that the denominators
become positive definite when we turn to Euclidean space.

In odd dimensions or with Dirac fermions $\psi $ we can just use Dirac-type
evanescent kinetic terms and write 
\begin{equation*}
S_{c\psi }+S_{\text{ev}\psi }=\int \bar{\psi}\left( i\gamma ^{\bar{\mu}}D_{%
\bar{\mu}}-m\right) \psi -\frac{\varsigma _{\psi }}{M}\int (\partial _{\hat{%
\mu}}\bar{\psi})(\partial ^{\hat{\mu}}\psi ),
\end{equation*}
where now $\varsigma _{\psi }$ is real. The propagator is 
\begin{equation*}
\frac{i\left( \bar{\slashed{p}}+m+\frac{\varsigma _{\psi }}{M}\hat{p}%
^{2}\right) }{\bar{p}^{2}-\left( m+\frac{\varsigma _{\psi }}{M}\hat{p}%
^{2}\right) ^{2}}.
\end{equation*}
Ultimately, the modified regularization we consider in this paper can always
be used in dimensions $>2$, and sometimes also in $d=2$.

\subsection{Scalars}

For definiteness, consider the theory 
\begin{equation}
S_{c\varphi }=\int (D_{\bar{\mu}}\varphi )^{\dagger }(D^{\bar{\mu}}\varphi
)-V(\varphi ^{\dagger }\varphi )+S_{Y},  \label{scfi}
\end{equation}
where $V$ is a potential and $S_{Y}$ denotes the Yukawa couplings and any
other types of perturbative corrections. The propagator of (\ref{scfi}) is $%
i/(\bar{p}^{2}-m^{2})$, therefore it does not behave correctly. We add the
quadratic evanescent terms 
\begin{equation}
S_{\text{ev}\varphi }=-\frac{\varsigma _{\varphi }}{M^{2}}\int (\hat{\partial%
}^{2}\varphi )^{\dagger }(\hat{\partial}^{2}\varphi )+\frac{\eta _{\varphi }%
}{M}\int (\partial _{\hat{\mu}}\varphi )^{\dagger }(\partial ^{\hat{\mu}%
}\varphi ).  \label{sevfi}
\end{equation}
The propagator of the total action $S_{c\varphi }+S_{\text{ev}\varphi }$ is
then 
\begin{equation*}
\frac{i}{D(\bar{p},\hat{p},m,\varsigma _{\varphi },\eta _{\varphi })}
\end{equation*}
and therefore has the right type of denominator. In some cases evanescent
vertices of the form 
\begin{equation}
\int (\hat{\partial}^{2}\varphi )^{\dagger }\varphi ^{2},\qquad \int (\hat{%
\partial}^{2}\varphi )^{\dagger }\varphi ^{3},  \label{noeva}
\end{equation}
etc., in $d=3,4$ might be allowed and should be added to the action. If no
parameters of negative weights are around, no other evanescent terms
compatible with weighted power counting and the global symmetries of the
theory can be constructed. Then $S_{c\varphi }+S_{\text{ev}\varphi }$ plus
the terms of type (\ref{noeva}) is the total regularized action.

\subsection{Gauge fields}

Now we switch to gauge fields. For definiteness, we take non-Abelian
Yang-Mills theory with a simple gauge group $G$, coupled to left-handed
fermions $\psi _{L}$. The gauge-invariant action is $S_{c}=S_{cA}+S_{c\psi }$%
, with 
\begin{equation*}
S_{cA}=-\frac{1}{4}\int F_{\bar{\mu}\bar{\nu}}^{a}F^{\bar{\mu}\bar{\nu}%
a},\qquad F_{\bar{\mu}\bar{\nu}}^{a}=\partial _{\bar{\mu}}A_{\bar{\nu}%
}^{a}-\partial _{\bar{\nu}}A_{\bar{\mu}}^{a}+gf^{abc}A_{\bar{\mu}}^{b}A_{%
\bar{\nu}}^{c}.
\end{equation*}

To keep track of gauge invariance during the renormalization algorithm, it
is convenient to use the Batalin-Vilkovisky formalism \cite{bata}. A set of
fields $\Phi ^{\alpha }=\{A_{\bar{\mu}}^{a},C^{a},\bar{C}^{a},B^{a},\psi
_{L},\bar{\psi}_{L}\}$ is defined, to collect the classical fields $\phi
=\{A_{\bar{\mu}}^{a},\psi _{L},\bar{\psi}_{L}\}$, the ghosts $C$, the
antighosts $\bar{C}$ and the Lagrange multipliers $B$ for the gauge fixing.
An external source $K_{\alpha }$ with opposite statistics is associated with
each $\Phi ^{\alpha }$, and coupled to the $\Phi ^{\alpha }$ symmetry
transformations $R^{\alpha }(\Phi )$. We have $K_{\alpha }=\{K^{\bar{\mu}%
a},K_{C}^{a},K_{\bar{C}}^{a},K_{B}^{a},K_{\psi },\bar{K}_{\psi }\}$. If $X$
and $Y$ are functionals of $\Phi $ and $K$ their \textit{antiparentheses}
are defined as 
\begin{equation}
(X,Y)\equiv \int \left( \frac{\delta _{r}X}{\delta \Phi ^{\alpha }}\frac{%
\delta _{l}Y}{\delta K_{\alpha }}-\frac{\delta _{r}X}{\delta K_{\alpha }}%
\frac{\delta _{l}Y}{\delta \Phi ^{\alpha }}\right) ,  \label{usa}
\end{equation}
where the integral is over spacetime points associated with repeated
indices. The \textit{master equation} $(S,S)=0$ must be solved for $S=S(\Phi
,K)$ in $D=d$ with the ``boundary condition'' $S=S_{c}$ at $C=\bar{C}%
=B=K_{\alpha }=0$. The solution 
\begin{equation}
S_{c}+S_{K}  \label{star}
\end{equation}
is the action we start with to quantize the theory, where 
\begin{eqnarray*}
S_{K}(\Phi ,K) &=&-\int R^{\alpha }(\Phi )K_{\alpha }=-\int (D_{\bar{\mu}%
}C^{a})K^{\bar{\mu}a}+\frac{g}{2}\int f^{abc}C^{b}C^{c}K_{C}^{a}-\int
B^{a}K_{\bar{C}}^{a} \\
&&+g\int \left( \bar{\psi}_{L}T^{a}C^{a}K_{\psi }+\bar{K}_{\psi
}T^{a}C^{a}\psi _{L}\right) ,
\end{eqnarray*}
encodes the symmetry transformations $R^{\alpha }(\Phi )$ of the fields, $D_{%
\bar{\mu}}C^{a}=\partial _{\bar{\mu}}C^{a}+gf^{abc}A_{\bar{\mu}}^{b}C^{c}$
being the covariant derivative of the ghosts. We can easily check that $%
(S_{c},S_{c})=(S_{c},S_{K})=(S_{K},S_{K})=0$ in arbitrary $D$ dimensions.

We gauge fix the theory with the gauge fermion 
\begin{equation*}
\Psi (\Phi )=\int \bar{C}^{a}\left( \partial ^{\bar{\mu}}A_{\bar{\mu}}^{a}+%
\frac{\xi }{2}B^{a}\right) ,
\end{equation*}
which means that we add 
\begin{equation*}
(S_{K},\Psi )=\int B^{a}\left( \partial ^{\bar{\mu}}A_{\bar{\mu}}^{a}+\frac{%
\xi }{2}B^{a}\right) -\int \bar{C}^{a}\partial ^{\bar{\mu}}D_{\bar{\mu}}C^{a}
\end{equation*}
to the action (\ref{star}), and obtain the gauge-fixed action 
\begin{equation}
S_{d}(\Phi ,K)=S_{c}+(S_{K},\Psi )+S_{K},  \label{sid}
\end{equation}
which still satisfies $(S_{d\hspace{0.01in}},S_{d\hspace{0.01in}})=0$
exactly in arbitrary $D$ dimensions.

As in the cases of fermions and scalars, the action $S_{d}$ is not well
regularized, since its propagators depend only on the $d$-dimensional
components of the momenta. Thus, we add evanescent corrections $S_{\text{ev}%
} $, guided by weighted power counting. We have learned from the treatment
of fermions that the weights of the fields and the physical components of
the momenta coincide with their dimensions in units of mass, and two
derivatives $\hat{\partial}$ count as one derivative $\bar{\partial}$.
Denoting weights with square brackets, we have 
\begin{equation*}
\lbrack A_{\bar{\mu}}^{a}]=[\bar{C}^{a}]=[C^{a}]=\frac{d}{2}-1,\qquad
[B^{a}]=\frac{d}{2},\qquad [\bar{\partial}]=1,\qquad [\hat{\partial}]=\frac{1%
}{2}.
\end{equation*}
Then we take 
\begin{eqnarray}
S_{\text{ev}A} &=&\frac{\varsigma _{A}}{2M^{2}}\int A_{\bar{\mu}}^{a}(\hat{%
\partial}^{2})^{2}A^{\bar{\mu}a}+\frac{\eta _{A}}{2M}\int A_{\bar{\mu}}^{a}(%
\hat{\partial}^{2})A^{\bar{\mu}a},  \label{seva} \\
S_{\text{ev}C} &=&-\frac{\varsigma _{C}}{M^{2}}\int \bar{C}^{a}(\hat{\partial%
}^{2})^{2}C^{a}-\frac{\eta _{C}}{M}\int \bar{C}^{a}(\hat{\partial}^{2})C^{a},
\label{sevc}
\end{eqnarray}
for gauge fields and ghosts, respectively. Note that no evanescent terms can
be constructed with $B^{a}$.

The final step is to include all of the other terms allowed by weighted
power counting and ghost number conservation. We can distinguish
nonevanescent additional terms $\Delta S_{c}$ and evanescent additional
terms $\Delta S_{\text{ev}}$. The total classical action 
\begin{equation*}
S_{c}=S_{cA}+S_{c\psi }+\Delta S_{c}
\end{equation*}
must be such that the gauge-fixed action $S_{d}$ defined by formula (\ref%
{sid}) still satisfies $(S_{d},S_{d})=0$ exactly in arbitrary $D$
dimensions. For example, if the theory is nonrenormalizable (like the
standard model coupled to quantum gravity, or a low-energy effective field
theory) $\Delta S_{c}$ collects infinitely many corrections of higher
dimensions (such as two scalar-two fermion vertices, Pauli terms,
four-fermion vertices, etc.). Recall that in $d>4$ all gauge theories are
nonrenormalizable.

The total evanescent action reads 
\begin{equation*}
S_{\text{ev}}=S_{\text{ev}A}+S_{\text{ev}\psi }+S_{\text{ev}C}+\Delta S_{%
\text{ev}}.
\end{equation*}
For example, in $d=3$ we have $\Delta S_{\text{ev}}=$ $\Delta S_{\text{ev}A}$%
, where 
\begin{equation*}
\Delta S_{\text{ev}A}=\frac{1}{M}\int \varepsilon ^{\bar{\mu}\bar{\nu}\bar{%
\rho}}\left( \varsigma _{A}^{\prime }A_{\bar{\mu}}^{a}(\hat{\partial}%
^{2})\partial _{\bar{\nu}}A_{\bar{\rho}}^{a}+\zeta _{A}gf^{abc}A_{\bar{\mu}%
}^{a}A_{\bar{\nu}}^{b}\hat{\partial}^{2}A_{\bar{\rho}}^{c}\right) ,
\end{equation*}
$\varsigma _{A}^{\prime }$ and $\zeta _{A}$ being constants.

Finally, the gauge-fixed regularized action of chiral gauge theories\ reads 
\begin{equation}
S(\Phi ,K)=S_{c}+(S_{K},\Psi )+S_{K}+S_{\text{ev}}=S_{d}+S_{\text{ev}}.
\label{defac}
\end{equation}
Since $(S_{d\hspace{0.01in}},S_{d\hspace{0.01in}})=0$, the action (\ref%
{defac}) satisfies the deformed master equation 
\begin{equation}
(S,S)=\mathcal{O}(\varepsilon ).  \label{master}
\end{equation}

To prove that the CD regularization is consistent, we need to focus on the
propagators, which contain only parameters of non-negative weights. Thus,
even when the theory is nonrenormalizable, it is enough to study the
subsector where the parameters of negative weights are switched off. In this
subsector no new terms are allowed besides those listed so far. In
particular, no evanescent terms depending on the sources $K$ can be
constructed. Moreover, $\Delta S_{\text{ev}}=0$ in $d>3$.

Collecting all pieces together (and switching off the parameters of negative
dimensions), we find 
\begin{eqnarray}
S(\Phi ,K) &=&-\frac{1}{4}\int F_{\bar{\mu}\bar{\nu}}^{a}F^{\bar{\mu}\bar{\nu%
}a}+\int B^{a}\left( \partial ^{\bar{\mu}}A_{\bar{\mu}}^{a}+\frac{\xi }{2}%
B^{a}\right) +\frac{1}{2}\int A_{\bar{\mu}}^{a}\frac{\hat{\partial}^{2}}{M}%
\left( \varsigma _{A}\frac{\hat{\partial}^{2}}{M}+\eta _{A}\right) A^{\bar{%
\mu}a}  \notag \\
&&+\int \bar{\psi}_{L}i\gamma ^{\bar{\mu}}D_{\bar{\mu}}\psi _{L}+\frac{i}{2M}%
\int \left( \varsigma _{\psi }\psi _{L}^{T}\mathcal{\tilde{C}}\hat{\partial}%
^{2}\psi _{L}-\varsigma _{\psi }^{*}\bar{\psi}_{L}\mathcal{\tilde{C}}\hat{%
\partial}^{2}\bar{\psi}_{L}^{T}\right)  \notag \\
&&-\int \bar{C}^{a}\left( \partial ^{\bar{\mu}}D_{\bar{\mu}}+\frac{\varsigma
_{C}(\hat{\partial}^{2})^{2}}{M^{2}}+\eta _{C}\frac{\hat{\partial}^{2}}{M}%
\right) C^{a}+\Delta S_{\text{ev}A}+S_{K}.  \label{schir}
\end{eqnarray}
Observe that the ghosts still decouple in the Abelian case. The ghost
propagators are 
\begin{equation}
\langle C^{a}(p)\hspace{0.01in}\bar{C}^{b}(-p)\rangle _{0}=\frac{i\delta
^{ab}}{D(\bar{p},\hat{p},0,\varsigma _{C},\eta _{C})}.  \label{ghp}
\end{equation}
Ignoring $\Delta S_{\text{ev}A}$ for a moment, the propagators of the
multiplet made of $A_{\bar{\mu}}^{a}$ and $B^{a}$ are 
\begin{eqnarray*}
\langle A_{\bar{\mu}}^{a}(p)\hspace{0.01in}A_{\bar{\nu}}^{b}(-p)\rangle _{0}
&=&\frac{-i\delta ^{ab}}{D(\bar{p},\hat{p},0,\varsigma _{A},\eta _{A})}%
\left( \eta _{\bar{\mu}\bar{\nu}}+\frac{(\xi -1)p_{\bar{\mu}}p_{\bar{\nu}}}{%
D(\bar{p},\hat{p},0,\xi \varsigma _{A},\xi \eta _{A})}\right) , \\
\langle A_{\bar{\mu}}^{a}(p)\hspace{0.01in}B^{b}(-p)\rangle _{0} &=&\frac{%
-p_{\bar{\mu}}\delta ^{ab}}{D(\bar{p},\hat{p},0,\xi \varsigma _{A},\xi \eta
_{A})},\qquad \langle B^{a}(p)\hspace{0.01in}B^{b}(-p)\rangle _{0}=i\delta
^{ab}\frac{\hat{p}^{2}}{M}\frac{\frac{\varsigma _{A}}{M}\hat{p}^{2}-\eta _{A}%
}{D(\bar{p},\hat{p},0,\xi \varsigma _{A},\xi \eta _{A})}.
\end{eqnarray*}
All of them have correct denominators and correct structures to ensure the
locality of counterterms, according to weighted power counting. Note that we
must keep $\xi \neq 0$, which means that the Landau gauge is not available
in the CD regularization.

When we switch $\Delta S_{\text{ev}A}$ on, in $d=3$, the propagators remain
regular. Indeed, towards the end of subsection 2.5 we prove a theorem
stating that if the propagators of some action $S$ are regular, they remain
regular when $S$ is extended by adding new terms compatible with weighted
power counting, multiplied by independent parameters.

\subsection{Gravity}

Now we move to quantum gravity. We recall that $x^{\bar{\mu}\hspace{0.01in}}$
are the coordinates of the physical portion of spacetime, and $x^{\hat{\mu}%
\hspace{0.01in}}$ are those of the evanescent portion. The metric tensor
depends on both, like every other field, but its nontrivial components are
just the usual $d$-dimensional ones $g_{\bar{\mu}\bar{\nu}}$. Precisely, in
the evanescent sector we take the flat-space metric $g_{\hat{\mu}\hat{\nu}%
}=\eta _{\hat{\mu}\hat{\nu}}$, while the off-diagonal components $g_{\bar{\mu%
}\hat{\nu}}$ identically vanish, so we have 
\begin{equation}
g_{\mu \nu }(\bar{x},\hat{x})=\left( 
\begin{tabular}{cc}
$g_{\bar{\mu}\bar{\nu}}(\bar{x},\hat{x})$ & $0$ \\ 
$0$ & $\eta _{\hat{\mu}\hat{\nu}}$%
\end{tabular}
\right) .  \label{form}
\end{equation}
General changes of coordinates affect only $x^{\bar{\mu}\hspace{0.01in}}$,
and leave $x^{\hat{\mu}\hspace{0.01in}}$ unmodified: 
\begin{equation}
x^{\bar{\mu}\hspace{0.01in}\prime }=\xi ^{\bar{\mu}}(\bar{x},\hat{x}),\qquad
x^{\hat{\mu}\hspace{0.01in}\prime }=x^{\hat{\mu}}.  \label{diffeo}
\end{equation}
Under these transformations $g_{\bar{\mu}\bar{\nu}}$ and $\partial _{\bar{\mu%
}}$ transform as usual, and $\mathrm{d}^{D}x\sqrt{|g|}$ is invariant, where $%
g$ is the determinant of $g_{\bar{\mu}\bar{\nu}}$. Actually, $g_{\hat{\mu}%
\hat{\nu}}$, $g_{\bar{\mu}\hat{\nu}}$ and $\hat{\partial}$ also change, so $%
g_{\mu \nu }$ does not keep the form (\ref{form}). However, this is not a
problem, since we do not require that the evanescent sector of the theory be
invariant under the general coordinate transformations (\ref{diffeo}).

We start from the ordinary $d$-dimensional Hilbert action 
\begin{equation}
S_{cG}=S_{\mathrm{H}}=-\frac{1}{2\kappa ^{2}}\int \mathrm{d}^{D}x\sqrt{|g(%
\bar{x},\hat{x})|}R(\bar{x},\hat{x}),  \label{scg}
\end{equation}
the constant $\kappa $ having the dimension of an energy to the power $%
(2-D)/2$. In $D$ dimensions covariant derivatives, as well as the
Christoffel symbols, the Riemann and Ricci tensors and the Ricci scalar are
defined by exactly the same formulas that hold in $d$ dimensions, therefore
they transform as usual: 
\begin{eqnarray*}
\Gamma _{\bar{\nu}\bar{\rho}}^{\bar{\mu}} &=&\frac{1}{2}g^{\bar{\mu}\bar{%
\sigma}}(\partial _{\bar{\nu}}g_{\bar{\rho}\bar{\sigma}}+\partial _{\bar{\rho%
}}g_{\bar{\nu}\bar{\sigma}}-\partial _{\bar{\sigma}}g_{\bar{\nu}\bar{\rho}%
}),\qquad R_{\hspace{0.02in}\hspace{0.02in}\hspace{0.02in}\bar{\nu}\bar{\rho}%
\bar{\sigma}}^{\bar{\mu}}=\partial _{\bar{\rho}}\Gamma _{\bar{\nu}\bar{\sigma%
}}^{\bar{\mu}}-\partial _{\bar{\sigma}}\Gamma _{\bar{\nu}\bar{\rho}}^{\bar{%
\mu}}+\Gamma _{\bar{\alpha}\bar{\rho}}^{\bar{\mu}}\Gamma _{\bar{\nu}\bar{%
\sigma}}^{\bar{\alpha}}-\Gamma _{\bar{\alpha}\bar{\sigma}}^{\bar{\mu}}\Gamma
_{\bar{\nu}\bar{\rho}}^{\bar{\alpha}}, \\
R_{\bar{\mu}\bar{\nu}} &=&R_{\hspace{0.02in}\hspace{0.02in}\hspace{0.02in}%
\bar{\mu}\bar{\rho}\bar{\nu}}^{\bar{\rho}},\qquad R=g^{\bar{\mu}\bar{\nu}}R_{%
\bar{\mu}\bar{\nu}}.\qquad
\end{eqnarray*}
Basically, the evanescent components $x^{\hat{\mu}}$ of the coordinates are
treated as external parameters. Clearly, (\ref{scg})\ is invariant under (%
\ref{diffeo}) in arbitrary $D$ dimensions.

Now we introduce the ghosts of diffeomorphisms $C^{\bar{\mu}}$, as well as
the antighosts $\bar{C}_{\bar{\mu}}$ and the Lagrange multipliers $B_{\bar{%
\mu}}$, and an external source $K$ for every field. All of them depend on $%
\bar{x},\hat{x}$. The infinitesimal gauge transformations are collected into
the functional 
\begin{equation}
S_{K}=\int (g_{\bar{\mu}\bar{\rho}}\partial _{\bar{\nu}}C^{\bar{\rho}}+g_{%
\bar{\nu}\bar{\rho}}\partial _{\bar{\mu}}C^{\bar{\rho}}+C^{\bar{\rho}%
}\partial _{\bar{\rho}}g_{\bar{\mu}\bar{\nu}})K^{\bar{\mu}\bar{\nu}}+\int C^{%
\bar{\rho}}(\partial _{\bar{\rho}}C^{\bar{\mu}})K_{\bar{\mu}}^{C}-\int B_{%
\bar{\mu}}K_{\bar{C}}^{\bar{\mu}}.  \label{SK}
\end{equation}
As before, it is easy to check that $%
(S_{cG},S_{cG})=(S_{cG},S_{K})=(S_{K},S_{K})=0$ in $D$ dimensions.

For several applications it is important to preserve invariance under 
\textit{rigid diffeomorphisms}, which are the coordinate transformations 
\begin{equation}
x^{\bar{\mu}\hspace{0.01in}\prime }=M_{\bar{\nu}}^{\bar{\mu}}x^{\bar{\nu}%
},\qquad x^{\hat{\mu}\hspace{0.01in}\prime }=x^{\hat{\mu}},  \label{genco}
\end{equation}
where $M_{\bar{\nu}}^{\bar{\mu}}$ is an arbitrary invertible constant
matrix. The action $S_{cG}$ is obviously invariant under rigid
diffeomorphisms, but also $S_{K}$ is, if we declare that the sources $K$
are, according to the case, scalar densities, vector densities or tensor
densities of weight 1. That means, in practice, that they carry a hidden $%
\sqrt{|g|}$.

To ensure invariance under rigid diffeomorphisms, it is sufficient to
express all fields and derivatives $\bar{\partial}$ using lower spacetime
indices, contract those indices with the inverse metric tensor $g^{\bar{\mu}%
\bar{\nu}} $ everywhere, and finally multiply by an appropriate power of $%
\sqrt{|g|}$, to obtain scalar densities of weight 1. Derivatives $\hat{%
\partial}$, instead, must be contracted with $\eta ^{\hat{\mu}\hat{\nu}}$,
to ensure $SO(-\varepsilon )$ invariance. Preserving invariance under rigid
diffeomorphisms is convenient for some applications (see section 5),
because, among other things, it constrains the forms of counterterms and
allows us to work without introducing ``second metrics''. By that we mean
any additional metrics (including the flat-space metric $\eta _{\bar{\mu}%
\bar{\nu}}$) that are often used for gauge-fixing and regularization
purposes.

We want to show that the CD regularization is fully compatible with
invariance under rigid diffeomorphisms. First, it is possible to choose
gauge-fixing conditions that preserve this global symmetry. For example, we
can take the gauge fermion 
\begin{equation}
\Psi =-\int \sqrt{|g|}\bar{C}_{\bar{\mu}}\left[ \frac{1}{2\kappa }(\partial
_{\bar{\nu}}g^{\bar{\mu}\bar{\nu}}+\lambda g^{\bar{\mu}\bar{\nu}}g_{\bar{\rho%
}\bar{\sigma}}\partial _{\bar{\nu}}g^{\bar{\rho}\bar{\sigma}})+\frac{\xi }{2}%
g^{\bar{\mu}\bar{\nu}}B_{\bar{\nu}}\right] ,  \label{gfer}
\end{equation}
where $\lambda $ and $\xi $ are gauge-fixing parameters. As usual, the
action is gauge fixed by adding $(S_{K},\Psi )$: 
\begin{equation}
S_{d}=S_{cG}+(S_{K},\Psi )+S_{K}.  \label{sdgf}
\end{equation}
We clearly have $(S_{d},S_{d})=0$ in $D$ dimensions.

At this point we observe that the action (\ref{sdgf}) is not equipped with
well-regularized propagators, so we must add evanescent terms consistent
with weighted power counting. From (\ref{sdgf}) we derive the weight
assignments 
\begin{equation*}
\lbrack g_{\bar{\mu}\bar{\nu}}]=[C^{\bar{\mu}}]=0,\qquad [\bar{C}_{\bar{\mu}%
}]=\frac{d}{2}-1,\qquad [B_{\bar{\mu}}]=\frac{d}{2},\qquad [K^{\bar{\mu}\bar{%
\nu}}]=[K_{\bar{\mu}}^{C}]=d-1,\qquad [K_{\bar{C}}^{\bar{\mu}}]=\frac{d}{2}.
\end{equation*}
We have used $[\Phi ^{\alpha }]+[K_{\alpha }]=d-1$ for every $\alpha $. We
also want to arrange the regularizing terms so that the full gauge-fixed
CD-regularized action is invariant under rigid diffeomorphisms.

We start adding the evanescent quadratic terms 
\begin{equation}
S_{\text{ev}G}=\frac{1}{8\kappa ^{2}M^{2}}\int \sqrt{|g|}\left( \varsigma
_{G}(\hat{\partial}^{2}g_{\bar{\mu}\bar{\nu}})(\hat{\partial}^{2}g^{\bar{\mu}%
\bar{\nu}})+\varsigma _{G}^{\prime }(g_{\bar{\alpha}\bar{\beta}}\hat{\partial%
}^{2}g^{\bar{\alpha}\bar{\beta}})^{2}\right) +\frac{\varsigma _{C_{G}}}{%
2\kappa M^{2}}\int \sqrt{|g|}\bar{C}_{\bar{\mu}}(\hat{\partial}^{2})^{2}C^{%
\bar{\mu}},  \label{sevg}
\end{equation}
which are the key ones to make propagators well behaved. Other evanescent
terms can be included in $S_{\text{ev}G}$, such as: ($i$)\ quadratic terms
similar to those of (\ref{sevg}), but with just two $\hat{\partial}$'s
instead of four, contracted in various ways, and ($ii$)\ terms that
contribute only to vertices when the metric tensor is expanded around flat
space, for example 
\begin{equation*}
\frac{1}{\kappa ^{4}M^{2}}\int \sqrt{|g|}(\partial _{\hat{\alpha}}g_{\bar{\mu%
}\bar{\nu}})(\partial ^{\hat{\alpha}}g^{\bar{\nu}\bar{\rho}})(\partial _{%
\hat{\beta}}g_{\bar{\rho}\bar{\sigma}})(\partial ^{\hat{\beta}}g^{\bar{\sigma%
}\bar{\mu}}).
\end{equation*}
The total gauge-fixed action is then 
\begin{equation}
S(\Phi ,K)=S_{cG}+(S_{K},\Psi )+S_{K}+S_{\text{ev}G}=S_{d}+S_{\text{ev}G},
\label{sfk}
\end{equation}
and is clearly such that $(S,S)=\mathcal{O}(\varepsilon )$. Indeed, the
nonevanescent part $S_{d}$ satisfies the master equation exactly, while the
evanescent part violates the master equation, since the derivatives $\hat{%
\partial}$ are noncovariant.

Now we show that the propagators of (\ref{sfk}) are indeed well behaved. We
expand around flat spacetime by writing 
\begin{equation}
g_{\bar{\mu}\bar{\nu}}=\eta _{\bar{\mu}\bar{\nu}}+2\kappa \phi _{\bar{\mu}%
\bar{\nu}},\qquad C^{\bar{\mu}}=\kappa \tilde{C}^{\bar{\mu}},  \label{espf}
\end{equation}
and work out the expansion to the quadratic order. From $S_{cG}$ we obtain 
\begin{equation}
\frac{1}{2}\int \left( (\partial _{\bar{\alpha}}\phi ^{\bar{\mu}\bar{\nu}%
})(\partial ^{\bar{\alpha}}\phi _{\bar{\mu}\bar{\nu}})-(\partial _{\bar{%
\alpha}}\phi )(\partial ^{\bar{\alpha}}\phi )-2(\partial _{\bar{\mu}}\phi ^{%
\bar{\mu}\bar{\nu}})(\partial ^{\bar{\rho}}\phi _{\bar{\rho}\bar{\nu}%
})-2\phi (\partial _{\bar{\mu}}\partial _{\bar{\nu}}\phi ^{\bar{\mu}\bar{\nu}%
})\right) ,  \label{pf}
\end{equation}
where $\phi =\phi _{\bar{\mu}}^{\bar{\mu}}$ and indices are raised and
lowered with the flat metric $\eta _{\bar{\mu}\bar{\nu}}$. The gauge fixing $%
(S_{K},\Psi )$ contributes with 
\begin{equation*}
\int B_{\bar{\mu}}\left( \partial _{\bar{\nu}}\phi ^{\bar{\mu}\bar{\nu}%
}+\lambda \partial ^{\bar{\mu}}\phi -\frac{\xi }{2}B^{\bar{\mu}}\right) +%
\frac{1}{2}\int \bar{C}_{\bar{\mu}}\left( \partial _{\bar{\nu}}\partial ^{%
\bar{\nu}}\tilde{C}^{\bar{\mu}}+(1+2\lambda )\partial ^{\bar{\mu}}\partial _{%
\bar{\nu}}\tilde{C}^{\bar{\nu}}\right) .
\end{equation*}
Finally, the quadratic part of the evanescent sector $S_{\text{ev}G}$ is 
\begin{equation*}
-\frac{1}{2M^{2}}\int \left( \varsigma _{G}\phi _{\bar{\mu}\bar{\nu}}(\hat{%
\partial}^{2})^{2}\phi ^{\bar{\mu}\bar{\nu}}-\varsigma _{G}^{\prime }\phi (%
\hat{\partial}^{2})^{2}\phi \right) +\frac{\varsigma _{C_{G}}}{2M^{2}}\int 
\bar{C}_{\bar{\mu}}(\hat{\partial}^{2})^{2}\tilde{C}^{\bar{\mu}}
\end{equation*}
plus similar terms obtained making the substitutions $(\hat{\partial}%
^{2})^{2}/M^{2}\rightarrow \hat{\partial}^{2}/M$ and $\varsigma \rightarrow
\eta $.

The propagators of the multiplet $\phi _{\bar{\mu}\bar{\nu}},B_{\bar{\rho}}$
are very involved. We have worked them out with the help of a computer
program. We do not give the result here, but just report that they have the
right structure to make the CD regularization work, as long as $\xi \neq 0$, 
$\lambda \neq -1$, $\varsigma _{G}\neq d\varsigma _{G}^{\prime }$ and $d>2$.
The denominators are polynomials $P_{2w}(\bar{p},\hat{p})$ of even weights $%
2w$ such that both monomials $(\bar{p}^{2})^{w}$ and $(\hat{p}^{2})^{2w}$
are multiplied by nonvanishing coefficients. Moreover, the propagators fall
off with the correct velocities in all directions of integration.

The ghost propagator is 
\begin{equation*}
\langle \tilde{C}^{\bar{\mu}}(p)\hspace{0.01in}\bar{C}_{\bar{\nu}%
}(-p)\rangle _{0}=-\frac{2i}{D(\bar{p},\hat{p},0,\varsigma _{C_{G}},\eta
_{C_{G}})}\left( \delta _{\bar{\nu}}^{\bar{\mu}}-\frac{(1+2\lambda )p^{\bar{%
\mu}}p_{\bar{\nu}}}{2(1+\lambda )D\left( \bar{p},\hat{p},0,\frac{\varsigma
_{C_{G}}}{2(1+\lambda )},\frac{\eta _{C_{G}}}{2(1+\lambda )}\right) }\right)
\end{equation*}
and also has the right structure.

When the cosmological constant $\Lambda $ is turned on, we must treat it
nonperturbatively, as if it were the squared mass of a bosonic particle. For
the purposes of renormalization, since counterterms are polynomial in $%
\Lambda $ we can still expand around flat space, although flat space is no
longer an extreme of the classical action.

\subsection{Gravity in the vielbein formalism}

When gravity is coupled to matter, the actions of chiral fermions, scalars
and gauge fields must be covariantized, possibly adding nonminimal terms.
The covariantization of nonevanescent terms, such as $S_{cA}$, proceeds as
in $d$ dimensions, while the covariantization of the evanescent corrections,
such as $S_{\text{ev}A}$, is made only with respect to rigid
diffeomorphisms. For example, (\ref{seva}) and (\ref{sevc}) become 
\begin{eqnarray*}
S_{\text{ev}A} &=&\frac{\varsigma _{A}}{2M^{2}}\int \sqrt{|g|}g^{\bar{\mu}%
\bar{\nu}}A_{\bar{\mu}}^{a}(\hat{\partial}^{2})^{2}A_{\bar{\nu}}^{a}+\frac{%
\eta _{A}}{2M}\int \sqrt{|g|}g^{\bar{\mu}\bar{\nu}}A_{\bar{\mu}}^{a}(\hat{%
\partial}^{2})A_{\bar{\nu}}^{a}, \\
S_{\text{ev}C} &=&-\frac{\varsigma _{C}}{M^{2}}\int \sqrt{|g|}\bar{C}^{a}(%
\hat{\partial}^{2})^{2}C^{a}-\frac{\eta _{C}}{M}\int \sqrt{|g|}\bar{C}^{a}(%
\hat{\partial}^{2})C^{a},
\end{eqnarray*}
respectively.

When fermions are present, we must switch to the vielbein formalism. Then it
is necessary to distinguish spacetime indices $\mu ,\nu ,\ldots $ from
Lorentz indices $a,b,\ldots $, and split both into bar indices and hat
indices. The vielbein has physical components $e_{\bar{\mu}}^{\bar{a}}$ and
evanescent components $e_{\bar{\mu}}^{\hat{a}}=e_{\hat{\mu}}^{\bar{a}}=0$
and $e_{\hat{\mu}}^{\hat{a}}=\delta _{\hat{\mu}}^{\hat{a}}$. The spin
connection $\omega _{\bar{\mu}}^{\bar{a}\bar{b}}$ and the Riemann and Ricci
curvature tensors $R_{\bar{\mu}\bar{\nu}}^{\bar{a}\bar{b}}$, $R_{\bar{\mu}}^{%
\bar{a}}$ in $D$ dimensions are defined by the same formulas that hold in $d$
dimensions, the evanescent components $x^{\hat{\mu}}$ of the coordinates
being treated as external parameters.

The starting classical action of gravity coupled to (left-handed) chiral
fermions is 
\begin{equation}
S_{cG}^{\prime }=S_{cG}+\int e\bar{\psi}_{L}ie_{\bar{a}}^{\bar{\mu}}\gamma ^{%
\bar{a}}D_{\bar{\mu}}\psi _{L},  \label{spg}
\end{equation}
where $e$ is the determinant of the vielbein and $D_{\bar{\mu}}$ is the
gravitational covariant derivative. The functional (\ref{SK}) is replaced by 
\begin{eqnarray}
S_{K}^{\prime } &=&\int (e_{\bar{\rho}}^{\bar{a}}\partial _{\bar{\mu}}C^{%
\bar{\rho}}+C^{\bar{\rho}}\partial _{\bar{\rho}}e_{\bar{\mu}}^{\bar{a}}+C^{%
\bar{a}\bar{b}}e_{\bar{\mu}\bar{b}})K_{\bar{a}}^{\bar{\mu}}+\int C^{\bar{\rho%
}}(\partial _{\bar{\rho}}C^{\bar{\mu}})K_{\bar{\mu}}^{C}  \notag \\
&&+\int (C^{\bar{a}\bar{c}}\eta _{\bar{c}\bar{d}}C^{\bar{d}\bar{b}}+C^{\bar{%
\rho}}\partial _{\bar{\rho}}C^{\bar{a}\bar{b}})K_{\bar{a}\bar{b}}^{C}-\int
B_{\bar{\mu}}K_{\bar{C}}^{\bar{\mu}}-\int B_{\bar{a}\bar{b}}K_{\bar{C}}^{%
\bar{a}\bar{b}}  \label{spk} \\
&&+\int C^{\bar{\rho}}(\partial _{\bar{\rho}}\bar{\psi}_{L})K_{\psi }-\frac{i%
}{4}\int \bar{\psi}_{L}\sigma ^{\bar{a}\bar{b}}C_{\bar{a}\bar{b}}K_{\psi
}+\int K_{\bar{\psi}}C^{\bar{\rho}}(\partial _{\bar{\rho}}\psi _{L})-\frac{i%
}{4}\int K_{\bar{\psi}}\sigma ^{\bar{a}\bar{b}}C_{\bar{a}\bar{b}}\psi _{L}, 
\notag
\end{eqnarray}
where $\sigma ^{\bar{a}\bar{b}}=i[\gamma ^{\bar{a}},\gamma ^{\bar{b}}]/2$
and $C^{\bar{a}\bar{b}}$ are the ghosts of local Lorentz symmetry.
Obviously, the identities $(S_{cG}^{\prime },S_{cG}^{\prime
})=(S_{cG}^{\prime },S_{K}^{\prime })=(S_{K}^{\prime },S_{K}^{\prime })=0$
hold in $D$ dimensions.

The gauge fermion must be corrected to include gauge-fixing conditions for
local Lorentz symmetry. The common symmetric condition $e_{\mu }^{a}=e_{\nu
}^{b}\eta _{b\mu }\eta ^{\nu a}$ cannot be used, since it violates
invariance under rigid diffeomorphisms. It is better to start from the less
common gauge-fixing condition $\partial ^{\bar{\mu}}\omega _{\bar{\mu}}^{%
\bar{a}\bar{b}}=0$, write it in a form that is compatible with rigid
diffeomorphisms, and then include every term allowed by weighted power
counting, ghost number conservation and invariance under rigid
diffeomorphisms.

The gauge-fixing sector and the evanescent sector must also include parity
violating terms constructed with the tensor $\varepsilon ^{a_{1}\cdots
a_{d}} $. Those terms are specific of every $d$ and in general introduce a
large number of new parameters. To prove that the propagators of gravity in
the vielbein formalism are well defined in arbitrary $d$, we proceed in two
steps. We first ignore the parity violating terms belonging to the
gravitational sector and prove that the propagators are well behaved in that
particular case. Later, we prove that they remain well behaved when the
parity violating terms are turned on.

From (\ref{spg}), (\ref{spk}) and the gauge-fixing condition, we find the
weight assignments of fields and sources, which are 
\begin{eqnarray*}
\lbrack e_{\bar{\mu}}^{\bar{a}}] &=&[C^{\bar{\mu}}]=0,\qquad [C^{\bar{a}\bar{%
b}}]=[\omega _{\bar{\mu}}^{\bar{a}\bar{b}}]=1,\qquad [\bar{C}_{\bar{\mu}%
}]=[B^{\bar{a}\bar{b}}]=\frac{d}{2}-1,\qquad [\bar{C}_{\bar{a}\bar{b}}]=%
\frac{d}{2}-2, \\
\lbrack B_{\bar{\mu}}] &=&\frac{d}{2},\qquad [K_{\bar{a}}^{\bar{\mu}}]=[K_{%
\bar{\mu}}^{C}]=d-1,\qquad [K_{\bar{a}\bar{b}}^{C}]=d-2,\qquad [K_{\bar{C}}^{%
\bar{\mu}}]=\frac{d}{2},\qquad [K_{\bar{C}}^{\bar{a}\bar{b}}]=\frac{d}{2}+1.
\end{eqnarray*}

The parity invariant sector of the new gauge fermion is equal to 
\begin{eqnarray}
\Psi ^{\prime } &=&\Psi +\frac{1}{\kappa }\int \bar{C}_{\bar{a}\bar{b}%
}\partial _{\bar{\mu}}\left( eg^{\bar{\mu}\bar{\nu}}e^{\bar{\rho}\bar{a}%
}\partial _{\bar{\nu}}e_{\bar{\rho}}^{\bar{b}}+\lambda _{1}ee^{\bar{\mu}\bar{%
a}}g^{\bar{\rho}\bar{\nu}}\partial _{\bar{\rho}}e_{\bar{\nu}}^{\bar{b}%
}+\lambda _{2}ee^{\bar{\mu}\bar{a}}e^{\bar{\nu}\bar{b}}e_{\bar{c}}^{\bar{\rho%
}}\partial _{\bar{\rho}}e_{\bar{\nu}}^{\bar{c}}\right)  \notag \\
&&+\lambda _{3}\int e\bar{C}_{\bar{\mu}}(e^{\bar{\nu}\bar{a}}\partial _{\bar{%
\nu}}e_{\bar{a}}^{\bar{\mu}}-e^{\bar{\mu}\bar{a}}\partial _{\bar{\nu}}e_{%
\bar{a}}^{\bar{\nu}})+\int ee^{\bar{\mu}\bar{a}}e^{\bar{\nu}\bar{b}}\left(
\lambda _{4}B_{\bar{\nu}}\partial _{\bar{\mu}}\bar{C}_{\bar{a}\bar{b}%
}+\lambda _{5}\bar{C}_{\bar{\mu}}\partial _{\bar{\nu}}B_{\bar{a}\bar{b}%
}\right)  \notag \\
&&+\frac{1}{2}\int \bar{C}_{\bar{a}\bar{b}}\left[ e\xi _{1}B^{\bar{a}\bar{b}%
}+\xi _{2}\partial _{\bar{\mu}}\left( eg^{\bar{\mu}\bar{\nu}}\partial _{\bar{%
\nu}}B^{\bar{a}\bar{b}}\right) +\xi _{3}\partial _{\bar{\mu}}\left( ee^{\bar{%
\mu}\bar{a}}e_{\bar{c}}^{\bar{\nu}}\partial _{\bar{\nu}}B^{\bar{b}\bar{c}%
}\right) \right] ,  \label{psip}
\end{eqnarray}
where $\Psi $ is the same as in formula (\ref{gfer}), plus terms that
contribute only to vertices in the expansion (\ref{expaf}).

The nonevanescent sector of the total gauge-fixed action is then 
\begin{equation*}
S_{d}^{\prime }(\Phi ,K)=S_{cG}^{\prime }+(S_{K}^{\prime },\Psi ^{\prime
})+S_{K}^{\prime },
\end{equation*}
and satisfies $(S_{d}^{\prime },S_{d}^{\prime })=0$ in $D$ dimensions.

The evanescent sector (\ref{sevg}) is turned into 
\begin{eqnarray}
S_{\text{ev}G}^{\prime } &=&S_{\text{ev}G}+\frac{i}{2M}\int e\left(
\varsigma _{\psi }\psi _{L}^{T}\mathcal{\tilde{C}}\hat{\partial}^{2}\psi
_{L}-\varsigma _{\psi }^{\ast }\bar{\psi}_{L}\mathcal{\tilde{C}}\hat{\partial%
}^{2}\bar{\psi}_{L}^{T}\right)  \notag \\
&&+\frac{\varsigma _{1}}{4\kappa ^{2}M^{2}}\int e(\hat{\partial}^{2}e_{\bar{%
\mu}\bar{a}})\left( \hat{\partial}^{2}e^{\bar{\mu}\bar{a}}-e_{\bar{\nu}}^{%
\bar{a}}e_{\bar{b}}^{\bar{\mu}}\hat{\partial}^{2}e^{\bar{\nu}\bar{b}}\right)
\notag \\
&&+\frac{\varsigma _{2}}{\kappa M^{2}}\int e(\hat{\partial}^{2}B^{\bar{a}%
\bar{b}})e_{\bar{a}}^{\bar{\mu}}(\hat{\partial}^{2}e_{\bar{\mu}\bar{b}})+%
\frac{\varsigma _{3}}{2M^{2}}\int e(\hat{\partial}^{2}B^{\bar{a}\bar{b}})(%
\hat{\partial}^{2}B_{\bar{a}\bar{b}})  \notag \\
&&-\frac{1}{\kappa M^{2}}\int e\bar{C}_{\bar{a}\bar{b}}(\hat{\partial}%
^{2})^{2}\left( \varsigma _{4}C^{\bar{a}\bar{b}}+\varsigma _{5}e^{\bar{\mu}%
\bar{a}}e_{\bar{\rho}}^{\bar{b}}\partial _{\bar{\mu}}C^{\bar{\rho}}\right) .
\label{sevgp}
\end{eqnarray}%
Again, we can also add\ evanescent quadratic terms with just two $\hat{%
\partial}$'s instead of four, and one power of $M$ in the denominator
instead of two, contracted in various ways, plus evanescent terms that
contribute only to vertices in the expansion (\ref{expaf}).

Finally, the total gauge-fixed action is 
\begin{equation*}
S^{\prime }(\Phi ,K)=S_{cG}^{\prime }+(S_{K}^{\prime },\Psi ^{\prime
})+S_{K}^{\prime }+S_{\text{ev}G}^{\prime }=S_{d}^{\prime }(\Phi ,K)+S_{%
\text{ev}G}^{\prime },
\end{equation*}
and satisfies $(S^{\prime },S^{\prime })=\mathcal{O}(\varepsilon )$.

To study the propagators, we expand around flat space by writing 
\begin{equation}
e_{\bar{\mu}}^{\bar{a}}=(\mathrm{e}^{\kappa \phi })_{\bar{\mu}}^{\bar{b}%
}\left( \mathrm{e}^{\kappa \chi }\right) _{\bar{b}}^{\bar{a}},\qquad e_{\bar{%
\mu}}^{\hat{a}}=e_{\hat{\mu}}^{\bar{a}}=0,\qquad e_{\hat{\mu}}^{\hat{a}%
}=\delta _{\hat{\mu}}^{\hat{a}},\qquad C^{\bar{\mu}}=\kappa \tilde{C}^{\bar{%
\mu}},\qquad C^{\bar{a}\bar{b}}=\kappa \tilde{C}^{\bar{a}\bar{b}},
\label{expaf}
\end{equation}
where $\phi $ and $\chi $ are matrices with entries $\phi _{\bar{\mu}}^{\bar{%
b}}$ and $\chi _{\bar{b}}^{\bar{a}}$, such that $\phi _{\bar{\mu}\bar{a}%
}\equiv \phi _{\bar{\mu}}^{\bar{b}}\eta _{\bar{b}\bar{a}}$ is symmetric and $%
\chi _{\bar{\mu}\bar{a}}\equiv \chi _{\bar{\mu}}^{\bar{b}}\eta _{\bar{b}\bar{%
a}}$ is antisymmetric. Then we concentrate on the terms that are quadratic
in the fields. We write the quadratic part of the gauge-fixed regularized
gravitational action in compact form as 
\begin{equation*}
\frac{1}{2}\int \phi _{i}\tilde{Q}^{ij}\phi _{j},
\end{equation*}
where $\phi _{i}=\{\phi _{\bar{\mu}\bar{a}},B_{\bar{\mu}},\chi _{\bar{\mu}%
\bar{a}},B^{\bar{a}\bar{b}}\}$ is a multiplet collecting the fluctuations $%
\phi _{\bar{\mu}\bar{a}}$ and $\chi _{\bar{\mu}\bar{a}}$ of the vielbein
around flat space, as well as the Lagrange multipliers $B_{\bar{\mu}}$ and $%
B^{\bar{a}\bar{b}}$. Switching to momentum space, $\tilde{Q}^{ij}$ turns
into a matrix $Q^{ij}$ whose entries depend polynomially on the momentum $p$
and the various parameters it contains.

The gravitational propagators $P_{ij}=i(Q^{-1})_{ij}$ are much more involved
than in the metric-tensor formalism. To simplify the proof that they are
indeed well behaved, we first establish a useful property. Let $\lambda $
denote a subset of the parameters contained in $Q$. If the decomposition 
\begin{equation*}
Q^{ij}=Q_{0}^{ij}+R_{\lambda }^{ij},
\end{equation*}
where $Q_{0}=\left. Q\right| _{\lambda =0}$, is such that $%
(P_{0})_{ij}=i(Q_{0}^{-1})_{ij}$ are well behaved, then $%
P_{ij}=i(Q^{-1})_{ij}$ are also well behaved.

To prove this fact, we define the parameters $\lambda $ so that $R_{\lambda
}^{ij}$ is a linear combination of terms multiplied by $\lambda $. Since $%
(Q_{0}^{-1})_{ij}$ exists, the eigenvalues of $Q_{0}$ are nonvanishing for
generic values of $p$ and the parameters contained in $Q_{0}$. Then, within
a certain nonvanishing radius of convergence for the parameters $\lambda $,
the eigenvalues of $Q_{0}^{-1}R_{\lambda }$ have absolute values smaller
than one, therefore the eigenvalues of $Q_{0}+R_{\lambda }$ are also
nonvanishing, the inverse of $Q_{0}+R_{\lambda }$ exists and the series 
\begin{equation}
\frac{1}{Q_{0}+R_{\lambda }}=P_{0}\sum_{n=0}^{\infty }(-1)^{n}(R_{\lambda
}P_{0})^{n}  \label{lhs}
\end{equation}%
is convergent. Write 
\begin{equation*}
\frac{1}{Q_{0}}=\frac{N_{0}}{\det Q_{0}},
\end{equation*}%
where $N_{0}$ is a polynomial matrix defined by this same equation. Within
the convergence radius, we also have 
\begin{equation}
\frac{1}{Q_{0}+R_{\lambda }}=\frac{N_{\lambda }}{\det (Q_{0}+R_{\lambda })},
\label{lhs2}
\end{equation}%
where $N_{\lambda }$ is a polynomial matrix and $N_{0}=\left. N\right\vert
_{\lambda =0}$. Formula (\ref{lhs2}) tells us that in the domain of
convergence the entries of $(Q+R_{\lambda })^{-1}$ are rational functions of 
$p$ and the parameters. But then formula (\ref{lhs2}) also holds outside the
domain of convergence, for generic values of $p$ and the parameters, because
the algebraic operations that give 
\begin{equation*}
\frac{\left( Q_{0}+R_{\lambda }\right) N_{\lambda }}{\det (Q_{0}+R_{\lambda
})}=1
\end{equation*}%
are exactly the same. Thus formula (\ref{lhs2})\ gives the propagators
whenever $p$ and the parameters have nonexceptional values.

Now we study the ultraviolet behaviors of (\ref{lhs2}). We can focus on the
denominator $\det (Q_{0}+R_{\lambda })$. When $\bar{p}$ and/or $\hat{p}$
tend to infinity the corrections due to $R_{\lambda }$ cannot ruin the
ultraviolet behavior due to $\det (Q_{0})$. To see this, let $(\bar{p}%
^{2})^{w}$ and $(\hat{p}^{2})^{2w}$ denote the dominant monomials of $\det
(Q_{0})$ for $\bar{p}\rightarrow \infty $ and $\hat{p}\rightarrow \infty $,
respectively. They are multiplied by nonvanishing coefficients, because, by
assumption, the denominators of $(Q_{0})^{-1}$ have dominant terms
multiplied by nonvanishing coefficients. Since propagators contain only
parameters of non-negative weights, the corrections brought by $R_{\lambda }$
can at most change the coefficients of the dominant terms $(\bar{p}^{2})^{w}$
and $(\hat{p}^{2})^{2w}$ inside $\det (Q)$, but not their powers, which are
still $w$ and $2w$. In other words, the dominant terms of the denominators
continue to have nonvanishing coefficients for generic values of the
parameters, so the propagators are well behaved.

Thanks to this result we do not need to work out the most general
propagators $Q^{-1}$. It is sufficient to identify a particular case $Q_{0}$
such that the inverse $Q_{0}^{-1}$ is well behaved. It is convenient to
choose $\lambda _{1}=\lambda _{2}=\lambda _{3}=\lambda _{4}=\lambda _{5}=\xi
_{3}=0$, in (\ref{psip}), $\varsigma _{5}=-\varsigma _{4}$ in (\ref{sevgp}),
and turn off the coefficients of all parity violating terms. Then, the
multiplet $\phi _{i}$ splits into the two submultiplets $\{\phi _{\bar{\mu}%
\bar{a}},B_{\bar{\nu}}\}$ and $\{\chi _{\bar{\mu}\bar{a}},B^{\bar{b}\bar{c}%
}\}$, in the sense that the matrix $Q_{0}$ becomes block-diagonal in those
submultiplets. The ghost action also diagonalizes in the pairs $\bar{C}_{%
\bar{\mu}}$-$C^{\bar{\nu}}$ and $\bar{C}_{\bar{a}\bar{b}}$-$C^{\prime 
\hspace{0.01in}\bar{c}\bar{d}}$, where 
\begin{equation*}
C^{\prime \hspace{0.01in}\bar{a}\bar{b}}=C^{\bar{a}\bar{b}}-\frac{1}{2}%
\left( e^{\bar{\mu}\bar{a}}e_{\bar{\nu}}^{\bar{b}}-e^{\bar{\mu}\bar{b}}e_{%
\bar{\nu}}^{\bar{a}}\right) \partial _{\bar{\mu}}C^{\bar{\nu}}.
\end{equation*}%
It is easy to check that the propagators of the Lorentz ghosts are well
behaved. Moreover, the propagators of the submultiplet $\{\phi _{\bar{\mu}%
\bar{a}},B_{\bar{\nu}}\}$ and those of the ghosts of diffeomorphisms are
also well behaved, because they coincide with the ones of the previous
subsection. It remains to study the propagators of the submultiplet $\{\chi
_{\bar{\mu}\bar{a}},B^{\bar{b}\bar{c}}\}$. This can be done immediately,
since the relevant quadratic part is just 
\begin{equation*}
-\frac{\varsigma _{1}}{2}\int \chi ^{\bar{a}\bar{b}}\frac{(\hat{\partial}%
^{2})^{2}}{M^{2}}\chi _{\bar{a}\bar{b}}+\int B^{\bar{a}\bar{b}}\left( \bar{%
\partial}^{2}\chi _{\bar{a}\bar{b}}+\varsigma _{2}\frac{(\hat{\partial}%
^{2})^{2}}{M^{2}}\chi _{\bar{a}\bar{b}}\right) +\frac{1}{2}\int B^{\bar{a}%
\bar{b}}\left( \xi _{2}\bar{\partial}^{2}B_{\bar{a}\bar{b}}+\varsigma _{3}%
\frac{(\hat{\partial}^{2})^{2}}{M^{2}}B_{\bar{a}\bar{b}}\right) ,
\end{equation*}%
plus terms that are subdominant in the ultraviolet limit.

Having shown that the propagators are well behaved in the particular case we
have identified, the missing parameters can be turned on using the property
proved above, so we conclude that the most general propagators are also well
behaved.

The parity violating terms can be included with the same procedure. We do
not list all of them here, because they are too many. We just mention that
they can appear in the gauge fermion, such as 
\begin{equation*}
\frac{1}{\kappa }\int \bar{C}_{\bar{a}\bar{b}}\varepsilon ^{\bar{a}\bar{b}%
\bar{c}\bar{d}}\partial _{\bar{\mu}}\left( eg^{\bar{\mu}\bar{\nu}}\omega _{%
\bar{\nu}\bar{c}\bar{d}}\right) ,\qquad \int \bar{C}_{\bar{a}\bar{b}%
}\varepsilon ^{\bar{a}\bar{b}\bar{c}\bar{d}}\partial ^{\bar{\mu}}\left(
e\partial _{\bar{\mu}}B_{\bar{c}\bar{d}}\right) ,
\end{equation*}%
and in the evanescent sector, such as 
\begin{equation*}
\frac{1}{\kappa ^{2}M^{2}}\int e(\hat{\partial}^{2}e_{\bar{\mu}\bar{a}%
})\varepsilon ^{\bar{a}\bar{b}\bar{c}\bar{d}}e_{\bar{b}}^{\bar{\mu}}e_{\bar{c%
}}^{\bar{\nu}}(\hat{\partial}^{2}e_{\bar{\nu}\bar{d}}),\quad \frac{1}{M^{2}}%
\int e(\hat{\partial}^{2}B_{\bar{a}\bar{b}})\varepsilon ^{\bar{a}\bar{b}\bar{%
c}\bar{d}}(\hat{\partial}^{2}B_{\bar{c}\bar{d}}),\quad \frac{1}{M^{2}}\int e(%
\hat{\partial}^{2}\bar{C}_{\bar{a}\bar{b}})\varepsilon ^{\bar{a}\bar{b}\bar{c%
}\bar{d}}(\hat{\partial}^{2}C_{\bar{c}\bar{d}}).
\end{equation*}

\subsection{Chern-Simons theories}

Parity violating theories in odd dimensions $d$ may contain Chern-Simons
terms, which are built with the tensor $\varepsilon ^{\mu _{1}\cdots \mu
_{d}}$. The dimensional regularization of such theories raises issues that
are in some respects similar to those raised by the matrix $\gamma _{5}$ in
four dimensions. We start from three dimensional Chern-Simons Yang-Mills
theories, where 
\begin{eqnarray*}
S_{cA} &=&\frac{1}{2}\int \varepsilon ^{\bar{\mu}\bar{\nu}\bar{\rho}}A_{\bar{%
\mu}}^{a}\left( \partial _{\bar{\nu}}A_{\bar{\rho}}^{a}+\frac{g}{3}f^{abc}A_{%
\bar{\nu}}^{b}A_{\bar{\rho}}^{c}\right) , \\
S_{K} &=&-\int (D_{\bar{\mu}}C^{a})K^{\bar{\mu}a}+\frac{g}{2}\int
f^{abc}C^{b}C^{c}K_{C}^{a}-\int B^{a}K_{\bar{C}}^{a}.
\end{eqnarray*}
We choose the gauge fermion 
\begin{equation*}
\Psi (\Phi )=\int \bar{C}^{a}\left( \partial ^{\bar{\mu}}A_{\bar{\mu}}^{a}+%
\frac{h(i\bar{\partial})}{2}B^{a}\right) ,
\end{equation*}
where $h(i\bar{\partial})$ is an unspecified derivative operator. The weight
assignments are 
\begin{equation*}
\lbrack A]=[C]=[B]=1,\qquad [\bar{C}]=0,\qquad [g]=0,\qquad
[K_{A}]=[K_{C}]=1,\qquad [K_{\bar{C}}]=2,\qquad [h]=1.
\end{equation*}
However, $[h]=1$ implies that $h$ is not a polynomial, so we are forced to
set $h=0$.

The evanescent terms we can add compatibly with weighted power counting are 
\begin{equation}
S_{\text{ev}}=-\frac{\varsigma _{A}}{2M}\int A_{\bar{\mu}}^{a}(\hat{\partial}%
^{2})A^{\bar{\mu}a}+S_{\text{ev}C}-\frac{\varsigma _{B}}{2M}\int B_{{}}^{a}%
\hat{\partial}^{2}B^{a},  \label{lt}
\end{equation}
where $S_{\text{ev}C}$ is still given by (\ref{sevc}). Note the last term,
which is crucial to make the propagators well behaved even if $h=0$. The
ghost propagators coincide with (\ref{ghp}), while the $A$ and $B$
propagators are 
\begin{eqnarray*}
\langle A_{\bar{\mu}}^{a}(p)\hspace{0.01in}A_{\bar{\nu}}^{b}(-p)\rangle _{0}
&=&\frac{\delta ^{ab}}{D(\bar{p},\hat{p},0,\varsigma _{A}^{2},0)}\left[
\varepsilon _{\bar{\mu}\bar{\rho}\bar{\nu}}p^{\bar{\rho}}-i\frac{\hat{p}^{2}%
}{M}\left( \varsigma _{A}\eta _{\bar{\mu}\bar{\nu}}+\frac{(\varsigma
_{B}-\varsigma _{A})p_{\bar{\mu}}p_{\bar{\nu}}}{D(\bar{p},\hat{p}%
,0,\varsigma _{A}\varsigma _{B},0)}\right) \right] , \\
\langle A_{\bar{\mu}}^{a}(p)\hspace{0.01in}B^{b}(-p)\rangle _{0} &=&-\frac{%
p_{\bar{\mu}}\delta ^{ab}}{D(\bar{p},\hat{p},0,\varsigma _{A}\varsigma
_{B},0)},\qquad \langle B^{a}(p)\hspace{0.01in}B^{b}(-p)\rangle
_{0}=-i\varsigma _{A}\frac{\hat{p}^{2}}{M}\frac{\delta ^{ab}}{D(\bar{p},\hat{%
p},0,\varsigma _{A}\varsigma _{B},0)}.
\end{eqnarray*}
We see that all of them fall off with the appropriate weights in the
ultraviolet limit, in all directions of integration, as long as $\varsigma
_{A}$ and $\varsigma _{B}$ do not vanish.

There is no difficulty in studying Chern-Simons--Maxwell theory along the
same lines, that is to say include the term $F_{\bar{\mu}\bar{\nu}}^{a}F^{a%
\bar{\mu}\bar{\nu}}$ in $S_{cA}$. Since the Maxwell term prevails over the
Chern-Simons one in the ultraviolet limit, this model works like the
Yang-Mills theories studied in subsection 2.3, as if the Chern-Simons term
were absent.

The coupling to matter is straightforward. Instead, the coupling to gravity
must be discussed in detail, because when parity is violated the
gravitational Chern-Simons term 
\begin{equation*}
S_{GCS}=\frac{1}{2\alpha ^{2}}\int \varepsilon ^{\bar{\mu}\bar{\nu}\bar{\rho}%
}\Gamma _{\bar{\mu}\bar{\beta}}^{\bar{\alpha}}\left( \partial _{\bar{\nu}%
}\Gamma _{\bar{\rho}\bar{\alpha}}^{\bar{\beta}}+\frac{2}{3}\Gamma _{\bar{\nu}%
\bar{\gamma}}^{\bar{\beta}}\Gamma _{\bar{\rho}\bar{\alpha}}^{\bar{\gamma}%
}\right)
\end{equation*}
must be added to the Hilbert action $S_{\mathrm{H}}$ of (\ref{scg}).

Due to the large number of terms involved, it is convenient to
block-diagonalize the propagators. We expand around flat space by writing $%
\alpha =\kappa m^{1/2}$ and 
\begin{equation*}
e_{\bar{\mu}}^{\bar{a}}=\mathrm{e}^{\kappa \phi }(\mathrm{e}^{\alpha \tilde{%
\phi}})_{\bar{\mu}}^{\bar{b}}\left( \mathrm{e}^{\kappa \chi }\right) _{\bar{b%
}}^{\bar{a}},\qquad e_{\bar{\mu}}^{\hat{a}}=e_{\hat{\mu}}^{\bar{a}}=0,\qquad
e_{\hat{\mu}}^{\hat{a}}=\delta _{\hat{\mu}}^{\hat{a}},\qquad C^{\bar{\mu}%
}=\alpha \tilde{C}^{\bar{\mu}},\qquad C^{\bar{a}\bar{b}}=\kappa \tilde{C}^{%
\bar{a}\bar{b}},
\end{equation*}
where $\tilde{\phi}$ and $\chi $ are matrices with entries $\tilde{\phi}_{%
\bar{\mu}}^{\bar{b}}$ and $\chi _{\bar{b}}^{\bar{a}}$, such that $\tilde{\phi%
}_{\bar{\mu}\bar{a}}\equiv \tilde{\phi}_{\bar{\mu}}^{\bar{b}}\eta _{\bar{b}%
\bar{a}}$ is symmetric and traceless and $\chi _{\bar{\mu}\bar{a}}=\chi _{%
\bar{\mu}}^{\bar{b}}\eta _{\bar{b}\bar{a}}$ is antisymmetric. Moreover, $m$
is a parameter of dimension 1 that must be treated nonperturbatively (in
this sense, it behaves like an ordinary mass). Since $S_{GCS}$ is
conformally invariant, it does not depend on the conformal factor $\mathrm{e}%
^{\kappa \phi }$. \ Moreover, both $S_{GCS}$ and $S_{\text{H}}$ obviously do
not depend on $\chi _{\bar{\mu}\bar{a}}$, since they do not contain the
vielbein, but just the metric tensor. Precisely, we have 
\begin{equation*}
S_{GCS}(\tilde{\phi},\alpha )=\frac{1}{\alpha ^{2}}S_{GCS}^{\prime }(\alpha 
\tilde{\phi}),\qquad S_{\text{H}}(\phi ,\tilde{\phi},\kappa ,m)=\frac{1}{%
\kappa ^{2}}S_{\text{H}}^{\prime }(\kappa \phi ,\alpha \tilde{\phi}),
\end{equation*}
which are perturbative expansions in powers of $\alpha $ and $\kappa $,
respectively.

To avoid unnecessary complications, we search for a special case where it is
simpler to prove that the propagators are well defined. Using the trick
explained in the previous subsection we know that when we turn on the other
parameters the propagators remain well defined.

Since the Chern-Simons term is higher derivative, we need a
higher-derivative gauge fixing for $\tilde{\phi}_{\bar{\mu}\bar{a}}$ to
obtain well-behaved propagators. It is convenient to make the new gauge
fermion independent of the conformal factor to the lowest order around flat
space, using 
\begin{equation}
\frac{1}{2\alpha }\left( \partial _{\bar{\nu}}g^{\bar{\mu}\bar{\nu}}-\frac{1%
}{3}g^{\bar{\mu}\bar{\nu}}g_{\bar{\rho}\bar{\sigma}}\partial _{\bar{\nu}}g^{%
\bar{\rho}\bar{\sigma}}\right) =-g^{\bar{\mu}\bar{\nu}}\partial _{\bar{\rho}}%
\tilde{\phi}_{\bar{\nu}}^{\bar{\rho}}.  \label{sas}
\end{equation}
Moreover, it is not necessary to include $B_{\bar{\mu}}$-dependent terms. In
the sector $\chi _{\bar{\mu}\bar{a}}$-$B^{\bar{b}\bar{c}}$ we can take the
gauge fermion $\Psi ^{\prime }-\Psi $ of formula (\ref{psip}) with $\lambda
_{1}=\lambda _{2}=\lambda _{3}=\lambda _{4}=\lambda _{5}=\xi _{3}=0$. We
thus have 
\begin{eqnarray*}
\Psi ^{\prime } &=&\frac{1}{2\alpha }\int e\bar{C}_{\bar{\mu}}g^{\bar{\alpha}%
\bar{\beta}}\bar{\partial}_{\bar{\alpha}}\bar{\partial}_{\bar{\beta}}\left(
\partial _{\bar{\nu}}g^{\bar{\mu}\bar{\nu}}-\frac{1}{3}g^{\bar{\mu}\bar{\nu}%
}g_{\bar{\rho}\bar{\sigma}}\partial _{\bar{\nu}}g^{\bar{\rho}\bar{\sigma}%
}\right) \\
&&-\frac{1}{\kappa }\int e(\partial _{\bar{\mu}}\bar{C}_{\bar{a}\bar{b}})g^{%
\bar{\mu}\bar{\nu}}e^{\bar{\rho}\bar{a}}\partial _{\bar{\nu}}e_{\bar{\rho}}^{%
\bar{b}}-\frac{\xi _{2}}{2}\int e(\partial _{\bar{\mu}}\bar{C}_{\bar{a}\bar{b%
}})g^{\bar{\mu}\bar{\nu}}\partial _{\bar{\nu}}B^{\bar{a}\bar{b}},
\end{eqnarray*}
plus subdominant terms. The weight assignments read 
\begin{eqnarray*}
\lbrack \tilde{\phi}_{\bar{\mu}\bar{a}}] &=&0,\qquad [\phi ]=[\chi _{\bar{\mu%
}\bar{a}}]=\frac{1}{2},\qquad [C^{\bar{\mu}}]=0,\qquad [C^{\bar{a}\bar{b}%
}]=1,\qquad [\bar{C}_{\bar{\mu}}]=-1, \\
\lbrack \bar{C}^{\bar{a}\bar{b}}] &=&-\frac{1}{2},\qquad [B_{\bar{\mu}%
}]=0,\qquad [B^{\bar{a}\bar{b}}]=\frac{1}{2},\qquad [\kappa ]=-\frac{1}{2}%
,\qquad [\alpha ]=0.
\end{eqnarray*}

Summing 
\begin{equation*}
S_{\text{H}}+S_{GCS}+(S_{K}^{\prime },\Psi ^{\prime })+S_{K}^{\prime }
\end{equation*}
and expanding around flat space, we find a quadratic part of the form 
\begin{equation*}
\int \phi \bar{\partial}^{2}\phi +\tilde{\phi}\bar{\partial}^{3}\tilde{\phi}%
+m^{1/2}\tilde{\phi}\bar{\partial}^{2}\phi +m\tilde{\phi}\bar{\partial}^{2}%
\tilde{\phi}+B\bar{\partial}^{3}\tilde{\phi}+B^{\prime }\bar{\partial}%
^{2}\chi +B^{\prime }\bar{\partial}^{2}B^{\prime }+\bar{C}\bar{\partial}^{4}%
\tilde{C}+\bar{C}^{\prime }\bar{\partial}^{2}\tilde{C}^{\prime },
\end{equation*}
where $B,\bar{C},\tilde{C}$ stand for $B_{\bar{\mu}},\bar{C}_{\bar{\mu}},%
\tilde{C}^{\bar{\mu}}$ and $B^{\prime },\bar{C}^{\prime },\tilde{C}^{\prime
} $ stand for $B_{\bar{a}\bar{b}},\bar{C}_{\bar{a}\bar{b}},\tilde{C}^{\bar{a}%
\bar{b}}$.

The kinetic terms diagonalize in the blocks $\{\tilde{\phi}_{\bar{\mu}\bar{a}%
},B_{\bar{\nu}}\}$, $\{\chi _{\bar{\mu}\bar{a}},B^{\bar{b}\bar{c}}\}$ and $%
\{\phi \}$ at $m=0$. Thus, it is convenient to switch $m$ off, prove that
the propagators are well defined in that case, and then use the trick of the
previous subsection to conclude that they remain well behaved when $m$ is
turned on again.

The evanescent kinetic terms can be arranged to preserve the diagonal
structure just outlined. In particular, we can separate the conformal factor 
$\phi $ from $\tilde{\phi}_{\bar{\mu}\bar{a}}$ using the formulas 
\begin{equation}
\frac{1}{6\kappa }g^{\bar{\mu}\bar{\nu}}\hat{\partial}g_{\bar{\mu}\bar{\nu}}=%
\hat{\partial}\phi ,\qquad \frac{1}{2\alpha }\left( \hat{\partial}g_{\bar{\mu%
}\bar{\nu}}-\frac{1}{3}g_{\bar{\mu}\bar{\nu}}g^{\bar{\alpha}\bar{\beta}}\hat{%
\partial}g_{\bar{\alpha}\bar{\beta}}\right) =\hat{\partial}\tilde{\phi}_{%
\bar{\mu}\bar{\nu}}.  \label{dhatf}
\end{equation}
At $m=0$ the conformal factor $\phi $ behaves as an ordinary scalar field,
so its regularized propagator is straightforward, the evanescent kinetic
terms being 
\begin{equation*}
S_{\text{ev}\phi }=-\frac{\varsigma _{\phi }}{2\kappa ^{2}M^{2}}\int \sqrt{%
|g|}(g^{\bar{\mu}\bar{\nu}}\hat{\partial}g_{\bar{\mu}\bar{\nu}})\hat{\partial%
}^{2}(g^{\bar{\alpha}\bar{\beta}}\hat{\partial}g_{\bar{\alpha}\bar{\beta}}).
\end{equation*}

The block $\{\tilde{\phi}_{\bar{\mu}\bar{a}},B_{\bar{\nu}}\}$ can be
regularized by means of the evanescent terms 
\begin{eqnarray*}
S_{\text{ev}\tilde{\phi}} &=&-\frac{1}{2M}\int \sqrt{|g|}(\hat{\partial} _{%
\hat{\tau}}\tilde{\phi}_{\bar{\mu}\bar{\nu}})g^{\bar{\mu}\bar{\rho}}\left( 
\bar{\varsigma}_{\tilde{\phi}}g^{\bar{\alpha}\bar{\beta}}\bar{\partial}_{%
\bar{\alpha}}\bar{\partial}_{\bar{\beta}}+\varsigma _{\tilde{\phi}}\frac{(%
\hat{\partial}^{2})^{2}}{M^{2}}\right) g^{\bar{\nu}\bar{\sigma}}(\hat{%
\partial} ^{\hat{\tau}}\tilde{\phi}_{\bar{\rho}\bar{\sigma}}) \\
&&-\varsigma \int \sqrt{|g|}B_{\bar{\mu}}\frac{(\hat{\partial}^{2})^{2}}{%
M^{2}}g^{\bar{\mu}\bar{\nu}}\bar{\partial} _{\bar{\rho}}\tilde{\phi}_{\bar{%
\nu}}^{\bar{\rho}}-\frac{1}{2}\int \sqrt{|g|}g^{\bar{\mu}\bar{\nu}}B_{\bar{%
\mu}}\frac{\hat{\partial}^{2}}{M}\left( \varsigma _{B}^{\prime }g^{\bar{%
\alpha}\bar{\beta}}\bar{\partial}_{\bar{\alpha}}\bar{\partial}_{\bar{\beta}%
}+\varsigma _{B}\frac{(\hat{\partial}^{2})^{2}}{M^{2}}\right) B_{\bar{\nu}},
\end{eqnarray*}
where $\partial _{\bar{\rho}}\tilde{\phi}_{\bar{\nu}}^{\bar{\rho}}$ and $%
\hat{\partial}\tilde{\phi}_{\bar{\mu}\bar{\nu}}$ are shortcuts for the
expressions of formulas (\ref{sas}) and (\ref{dhatf}). The propagators of
this block are too involved to be reported here, but we have checked that
they are well behaved by means of a computer program.

The propagators of the block $\{\chi _{\bar{\mu}\bar{a}},B^{\bar{b}\bar{c}%
}\} $ and those of the Lorentz ghosts coincide with the ones studied in the
previous subsection, the evanescent terms being the last three lines of
formula (\ref{sevgp}). Note that we can set $\varsigma _{5}=0$ at $m=0$.

Finally, to make the propagators of $\langle C^{\bar{\mu}}(p)\bar{C}_{\bar{%
\nu}}(-p)\rangle _{0}$ well behaved it is sufficient to add the evanescent
kinetic terms 
\begin{equation*}
-\frac{\varsigma _{C}}{2\alpha }\int \sqrt{|g|}\bar{C}_{\bar{\mu}}\frac{(%
\hat{\partial}^{2})^{2}}{M^{2}}\left( 2g^{\bar{\alpha}\bar{\beta}}\bar{%
\partial}_{\bar{\alpha}}\bar{\partial}_{\bar{\beta}}+\varsigma _{C}\frac{(%
\hat{\partial}^{2})^{2}}{M^{2}}\right) C^{\bar{\mu}}
\end{equation*}

Since the propagators are well behaved in the particular case just examined,
we know that they are also well behaved when we turn on the missing terms,
therefore we conclude that they are well behaved in the most general case.

\section{Weighted power counting, locality of counterterms and
renormalization}

\setcounter{equation}{0}

Because propagators have the form (\ref{propag}), the locality of
counterterms and renormalization are controlled by weighted power counting 
\cite{halat}, instead of ordinary power counting. Weighted power counting
was introduced for Lorentz violating theories, where quadratic terms contain
the usual numbers of time derivatives (two for bosons, one for fermions, in
unitary theories), but are allowed to contain higher-space derivatives. So
far, we have only considered theories that are Lorentz symmetric in the
physical spacetime $\mathbb{R}^{d}$, but our treatment can be easily
generalized to include models where Lorentz symmetry is explicitly violated
in $\mathbb{R}^{d}$. What is important for our discussion is that Lorentz
symmetry is certainly violated in the continued spacetime $\mathbb{R}^{D}=%
\mathbb{R}^{d}\times \mathbb{R}^{-\varepsilon }$, so the results of refs. 
\cite{halat,lvsm} apply to our case. For comparison with those references,
it may be useful to take into account that the scale $M$ appearing here
plays the role of the ``scale of Lorentz violation'' $\Lambda _{L}$
appearing there. In this section we explain how weighted counting works
within the CD regularization.

The denominators of (\ref{propag})\ have dominant powers $(\bar{p}^{2})^{w}$
and $(\hat{p}^{2})^{2w}$ for $\bar{p}\rightarrow \infty $ and $\hat{p}%
\rightarrow \infty $, respectively. This tells us that $\bar{p}^{2}$ and $(%
\hat{p}^{2})^{2}$ are equally important in the ultraviolet limit. Weights of
fields, momenta and parameters have to be assigned so that and $\bar{p}^{2}$
and $(\hat{p}^{2})^{2}$ have the same weights, and the action $S$ and the
scale $M$ are weightless. For convenience, we take the energy to have weight
equal to 1, which coincides with its dimension in units of mass. Then $\bar{p%
}$ and $\hat{p}$ have weights $1$ and $1/2$, respectively, and the
polynomials (\ref{denni}) have weight equal to $2$. The weights of fields
and parameters then follow from the requirement that $M$ and the action be
weightless. Note that for the purposes of renormalization what are important
are the values of the weights (and dimensions) at $\varepsilon =0$, so we
define them as such.

Denoting weights with square brackets, we find 
\begin{eqnarray*}
\lbrack \bar{\partial}] &=&1,\qquad [\hat{\partial}]=\frac{1}{2},\qquad [%
\hat{x}]=-1,\qquad [\hat{x}]=-\frac{1}{2},\qquad [M]=0, \\
\lbrack \Phi ] &=&\frac{d-N_{\Phi }}{2},\qquad [gA_{\bar{\mu}}]=1,\qquad [g_{%
\bar{\mu}\bar{\nu}}]=0,
\end{eqnarray*}
where $N_{\Phi }$ are the numbers of derivatives $\bar{\partial}$ contained
in the dominant quadratic terms (\ref{dom}) of the fields $\Phi $. In the
ultraviolet limit the propagator of each field $\Phi $ must have the form (%
\ref{propag}) and fall off with weight $N_{\Phi }$, at least, in all
directions of integration.

The weights of fields, sources and derivatives $\bar{\partial}$ coincide
with their dimensions in units of mass. The weights of $\hat{\partial}$, $%
\hat{x}$ and $M$ are different from their dimensions in units of mass.
However, the combinations $M^{-1/2}\hat{\partial}$ and $M^{1/2}\hat{x}$ have
weights equal to their dimensions. Local actions are sums of spacetime
integrals of monomials constructed with the fields, the sources, and their
derivatives $\bar{\partial}$ and $\hat{\partial}$. If $n_{2}$ denotes the
number of derivatives $\hat{\partial}$, we write the coefficient in front of
each monomial as $\lambda /M^{n_{2}/2}$, to factorize an appropriate power
of $M$. Symbolically, the monomial reads 
\begin{equation}
\lambda \int \mathrm{d}^{D}x\hspace{0.01in}\bar{\partial}^{n_{1}}\left(
M^{-1/2}\hat{\partial}\right) ^{n_{2}^{{}}}\Phi ^{n_{3}}K^{n_{4}}.
\label{cunto}
\end{equation}
Using this convention, which we have tacitly adopted in the previous section
and maintain throughout the paper, all parameters $\lambda $ have weights
equal to their dimensions in units of mass.

At this point, we need to recall a few facts about the dimensional
regularization. Divergences are poles in $\varepsilon $, but the terms that
disappear when $D\rightarrow d$, called ``evanescences'' can be of two
types: \textit{formal} evanescences or \textit{analytic} evanescences.
Analytically evanescent terms, briefly denoted as ``aev'', are those that
factorize at least one $\varepsilon $, such as $\varepsilon F_{\bar{\mu}\bar{%
\nu}}F^{\bar{\mu}\bar{\nu}}$, $\varepsilon \bar{\psi}_{L}i\slashed{D}\psi
_{L}$, etc. Formally evanescent terms, briefly denoted as ``fev'', are those
that formally disappear when $D\rightarrow d$, but do not factorize powers
of $\varepsilon $. An example is $\psi _{L}^{T}\hat{\partial}^{2}\psi _{L}$.
Because the fields have no evanescent components, in the CD regularization
there are fewer formal evanescences than usual. They are built with the
tensor $\eta _{\hat{\mu}\hat{\nu}}$ and the extra components of the
coordinates $\hat{x}$, momenta $\hat{p}$ and derivatives $\hat{\partial}$.
The poles in $\varepsilon $ can multiply either nonevanescent terms or
formally evanescent terms. In the latter case we speak of \textit{divergent
evanescences}, also denoted as ``divev''. An example is $\psi _{L}^{T}\hat{%
\partial}^{2}\psi _{L}/\varepsilon $.

When we differentiate propagators with respect to any components of momenta,
their ultraviolet behaviors improve by an amount equal to the weight of the
derivatives. In particular, each derivative $\partial /\partial \bar{p}$
lowers the weight of the ultraviolet behavior by one unit, in all directions
of integration, and each derivative $\partial /\partial \hat{p}$ lowers the
weight by $1/2$. For example, 
\begin{equation}
\frac{\partial }{\partial p^{\bar{\mu}}}\frac{1}{D(\bar{p},\hat{p},\varsigma
)}=-\frac{2p^{\bar{\mu}}}{D(\bar{p},\hat{p},\varsigma )^{2}},\qquad \frac{%
\partial }{\partial p^{\hat{\mu}}}\frac{1}{D(\bar{p},\hat{p},\varsigma )}=%
\frac{4\varsigma \hat{p}^{2}p^{\hat{\mu}}}{M^{2}D(\bar{p},\hat{p},\varsigma
)^{2}}.  \label{dera}
\end{equation}
When we differentiate a diagram $G$ a sufficient number of times with
respect to any components $\bar{k}$, $\hat{k}$ of its external momenta $k$,
we obtain an overall convergent integral.

It is convenient to subtract away divergent evanescences like any other
divergences. Indeed, it is simple to show that this prescription ensures the
locality of all types of divergences and is consistent to all orders.
Consider a diagram $G$, subtract its subdivergences according this rule, and
call $G_{\text{sub}}$ the subtracted diagram. When $G_{\text{sub}}$ is
differentiated with respect to any components $\bar{k}$, $\hat{k}$ of its
external momenta, a completely convergent integral is obtained. Thus, both
the nonevanescent and formally evanescent divergences of $G_{\text{sub}}$
are local, and can be subtracted away. Iterating this procedure, both types
of divergences are local to arbitrarily high orders.

In the end, locality ensures that the divergences of $G_{\text{sub}}$ are
polynomials in $\bar{k}$ and $\hat{k}$. The weight $\omega (G)$ of each such
polynomial is equal to the weight of $G$ minus the weights of the parameters 
$\lambda $ that multiply the vertices belonging to $G$. Since the weight of
the loop integration measure $\mathrm{d}^{D}p=$ $\mathrm{d}^{d}\bar{p}%
\hspace{0.01in}\mathrm{d}^{-\varepsilon }\hat{p}$ is equal to $d$ (at $%
\varepsilon =0$), the overall degree of divergence $\omega (G)$ coincides
with the dimension of the integral associated with it.

All counterterms have the structure (\ref{cunto}) with $\lambda $ replaced
by a product of $\lambda $'s. The generating functional $\Gamma $ of
one-particle irreducible correlation functions is nonlocal and can be
expressed using only the quantities $\Phi $, $K$, $\bar{x}$, $M^{1/2}\hat{x}$%
, $\bar{\partial}$, $M^{-1/2}\hat{\partial}$ and $\lambda $, which have
weights equal to their dimensions. Symbolically, we write 
\begin{equation}
\Gamma =\Gamma (\Phi ,K,\bar{x},M^{1/2}\hat{x},\bar{\partial},M^{-1/2}\hat{%
\partial},\lambda ).  \label{cunto2}
\end{equation}
This result shows that the scale $M$ is always attached to the formally
evanescent quantities $\hat{x}$ and $\hat{\partial}$. The $\Gamma $
functional keeps the structure (\ref{cunto2}) throughout the renormalization
algorithm.

When divergences are removed, it is possible to take the limit $\varepsilon
\rightarrow 0$. This is done by first letting $\varepsilon $ tend to $0$
inside correlation functions, without affecting the formally evanescent
quantities $M^{1/2}\hat{x}$ and $M^{-1/2}\hat{\partial}$, and then dropping
the formally evanescent quantities. The combination of the two operations
gives the physical correlation functions. Formula (\ref{cunto2}) shows that
the scale $M$ drops out of physical correlation functions. Thanks to
weighted power counting, $M$ does not propagate into the physical sector of
the theory, and there is no need to take the limit $M\rightarrow \infty $.

The settings of the previous section show that all the requirements of
weighted power counting, in particular those concerning the correct
behaviors of propagators, can be satisfied in perturbatively unitary quantum
fields theories, where bosons have $N_{\Phi }=2$ and fermions have $N_{\Phi
}=1$. The CD regularization can be generalized to nonunitary theories, in
particular higher-derivative theories, where $N_{\Phi }$ can exceed those
values. For example, we may consider higher-derivative gravity \cite{stelle}
or even nonlocal theories \cite{modesto} coupled with chiral fermions. We do
not give details here, because the generalization is straightforward.

The parameters of positive, vanishing and negative weights are
superrenormalizable, strictly renormalizable and nonrenormalizable by
weighted power counting, respectively. The action of weighted power counting
renormalizable theories contains all the terms, and only those, that are
compatible with the nonanomalous symmetries and are multiplied by parameters 
$\lambda $ of non-negative weights.

If we ignore symmetries for the moment, all monomials must be multiplied by
independent parameters. Then, from the strict point of view of weighted
power counting, the counterterms have the same forms as the terms of the
classical action, and can be subtracted away by redefining parameters and
making field redefinitions.

When gauge and global symmetries are present, on the other hand, the
coefficients of various monomials are related to one another, therefore it
is necessary to verify which symmetries are nonanomalous and drop those that
are anomalous. In general gauge theories, the nonevanescent sector $S_{d}$
of the action $S$ must be exactly gauge invariant, therefore it includes the
usual terms and satisfies the master equation $(S_{d},S_{d})=0$. Instead,
the total CD-regularized action $S=$ $S_{d}+S_{\text{ev}}$ satisfies the
deformed master equation 
\begin{equation}
(S,S)=\mathcal{O}(\varepsilon ),  \label{dema}
\end{equation}
where the right-hand side collects both analytically and formally evanescent
terms.

Assuming that gauge symmetries are nonanomalous, namely that there exists a
subtraction scheme where anomalies cancel to all orders in perturbation
theory, we still have the problem of \textit{parametric completeness}. A
classical action is parametrically complete if it can be renormalized by
means of canonical transformations and parameter redefinitions. One way to
ensure that this is true is to start from the most general solution of (\ref%
{dema}). When the theory is renormalizable, that solution contains a finite
number of independent terms, therefore it is possible to work it out by
means of a direct analysis. When the theory is not renormalizable, it
contains an infinite number of independent terms, and the issue is more
involved. Moreover, in principle counterterms may deform the gauge symmetry
in observable ways and affect the classification of invariants.

In nonrenormalizable theories that are manifestly free of gauge anomalies
the problem of parametric completeness can be solved in various ways. If the
gauge symmetries satisfy certain linearity assumptions (satisfied among
others by Yang-Mills gauge symmetries, general covariance and local Lorentz
symmetry), it is convenient to use the background field method \cite{backc}.
Cohomological classifications of invariants \cite{coho} may also solve the
problem, if the result is of a suitable form. More generally, the classical
action can be algorithmically extended by brute force till it becomes
parametrically complete \cite{regnocoho}. The CD\ regularization may be
useful to prove these and related results in more economic ways, and
generalize them to (possibly nonrenormalizable) theories that are not
manifestly free of gauge anomalies (see section 5 for more details).


Obviously, the action (\ref{sfk}) of quantum gravity is not parametrically
complete. The classical action $S_{cG}$ must be extended to include all the\
nonevanescent counterterms generated by renormalization, which are
infinitely many, multiplied by new independent parameters. A convenient form
of the extended nonevanescent action $S_{c\text{ext}}$ is the one given in
ref. \cite{newQG}, where invariants are organized in an economic way by
means of field redefinitions. In particular, $S_{c\text{ext}}$ does not need
to contain higher-derivative quadratic corrections, since counterterms of
that type can be subtracted away by means of canonical transformations. This
property ensures that a perturbatively unitary theory is not driven by
renormalization into a higher-derivative, perturbatively nonunitary theory 
\cite{absence}.

The total extended action $S_{\text{ext}}=S_{c\text{ext}}+$ $S_{\text{ev}%
\hspace{0.01in}\text{ext}}$ must also include an extended version $S_{\text{%
ev}\hspace{0.01in}\text{ext}}$ of the evanescent sector $S_{\text{ev}}$,
which collects evanescent terms of higher weights compatible with invariance
under rigid diffeomorphisms and the other global nonanomalous symmetries.
The easiest way to classify the infinitely many terms contained in $S_{\text{%
ext}}$ is to organize the $\kappa $ dependence so that 
\begin{equation}
S_{\text{ext}}(\Phi ,K,\kappa ,\lambda )=\frac{1}{\kappa ^{2}}S_{\text{ext}%
}^{\prime }(\kappa \Phi ,\kappa K,\lambda ),  \label{esso}
\end{equation}
where $\lambda $ denotes any other parameters. Then radiative corrections
have the structure (see for example \cite{ABrenoYMLR}) 
\begin{equation*}
\Gamma _{\text{ext}}(\Phi ,K,\kappa ,\lambda )=\sum_{L=0}^{\infty }\kappa
^{2L-2}\Gamma _{\text{ext}L}^{\prime }(\kappa \Phi ,\kappa K,\lambda )
\end{equation*}
where $L$ labels the $L$-loop contributions.

The extensions just advocated do not affect the theory we perturb around
(which is provided by $S_{cG}+$ $S_{\text{ev}G}$ in the case of pure quantum
gravity). Indeed, the additional parameters $\varsigma _{\text{ext}}$
contained in $S_{c\text{ext}}+S_{\text{ev}\hspace{0.01in}\text{ext}}$ have
negative weights, therefore they are treated perturbatively. In this
respect, observe that the set of $\varsigma _{\text{ext}}$s includes the
coefficients of quadratic terms of high weights, such as 
\begin{equation}
\frac{\varsigma _{nG}}{2\kappa ^{2}M^{2}}\int \sqrt{|g|}(\hat{\partial}%
^{2}g_{\bar{\mu}\bar{\nu}})(g^{\bar{\alpha}\bar{\beta}}\bar{\partial}_{\bar{%
\alpha}}\bar{\partial}_{\bar{\beta}})^{n}(\hat{\partial}^{2}g^{\bar{\mu}\bar{%
\nu}}),  \label{twol}
\end{equation}
which in principle modify the propagators. However, since terms like this
are just introduced for regularization purposes, and do not affect the
physical poles, it is convenient to treat the quadratic contributions coming
from the extension $S_{c\text{ext}}+S_{\text{ev}\hspace{0.01in}\text{ext}}$
as ``two-leg vertices'' and drop all parameters of negative weights from
propagators, as we did in the previous section.

At the practical level, a nonrenormalizable theory must be truncated. The
truncation must contain finitely many terms and must correspond to some
perturbative expansion. Commonly, the truncated action $S_{T}$ of quantum
gravity contains the terms that have dimensions smaller than some $T$ in
units of mass. The perturbative expansion is then an expansion in powers of
the energy divided by some reference mass scale, typically the Planck mass.
In the CD regularization it is sufficient to truncate to the terms that have 
\textit{weights} smaller than $T$. The truncated action $S_{T}$ must solve
the master equation (\ref{master}) up to corrections that fall outside the
truncation. To build $S_{T}$, we can list all monomials that have weights
smaller than $T$, multiply them by independent parameters, and finally
relate the parameters to one another by imposing (\ref{master}), as well as
invariance under rigid diffeomorphisms and the other global nonanomalous
symmetries.

\section{One-loop chiral anomalies}

\setcounter{equation}{0}

In this section we use the CD-regularization technique to calculate the
known one-loop anomalies of chiral gauge theories in four dimensions. We
recall that the anomaly functional $\mathcal{A}$ is defined as the
antiparenthesis $(\Gamma ,\Gamma )$, which is also equal to the average $%
\langle (S,S)\rangle $. This identity can be proved making the change of
variables $\Phi ^{\alpha }\rightarrow \Phi ^{\alpha }+\theta (S,\Phi
^{\alpha })$ in the functional integral 
\begin{equation*}
Z(J,K)=\int [\mathrm{d}\Phi ]\exp \left( iS(\Phi ,K)+i\int \Phi ^{\alpha
}J_{\alpha }\right) ,
\end{equation*}
where $\theta $ is a constant anticommuting parameter, and using the fact
that the Jacobian determinant is equal to one in the (ordinary, as well as
chiral) dimensional regularization (for details, see for example \cite{backc}%
). Moreover, $\mathcal{A}$ satisfies the Wess-Zumino consistency conditions 
\cite{wesszumino}, which are expressed by the identity $(\Gamma ,\mathcal{A}%
)=0$ using the Batalin-Vilkovisky formalism.

We begin with chiral QED. The action is (\ref{schir}) with $\mathcal{C}%
=-i\gamma ^{0}\gamma ^{2}$, $T^{a}\rightarrow i$ and $\Delta S_{\text{ev}%
A}=0 $. We have 
\begin{eqnarray*}
(S,S) &=&2(S_{K},S_{\text{ev}A}+S_{\text{ev}\psi }+S_{\text{ev}C}) \\
&=&\frac{2g}{M}\int C(\varsigma _{\psi }\psi _{L}^{\alpha }\varepsilon
_{\alpha \beta }\hat{\partial}^{2}\psi _{L}^{\beta }-\varsigma _{\psi
}^{\ast }\psi _{L}^{\ast \alpha }\varepsilon _{\alpha \beta }\hat{\partial}%
^{2}\psi _{L}^{\ast \beta })+a \\
&=&2\int C(\partial _{\bar{\mu}}J^{\bar{\mu}})+2ig\int C\left( \psi
_{L}^{\ast }\frac{\delta _{l}\bar{S}}{\delta \psi _{L}^{\ast }}-\frac{\delta
_{r}\bar{S}}{\delta \psi _{L}}\psi _{L}\right) +a,
\end{eqnarray*}%
where $J^{\bar{\mu}}=g\psi _{L}^{\dagger }\bar{\sigma}^{\bar{\mu}}\psi _{L}$
is the gauge current, $\bar{S}(\Phi )=S(\Phi ,0)$ and 
\begin{equation*}
a=2\int A_{\bar{\mu}}\frac{\hat{\partial}^{2}}{M}\left( \varsigma _{A}\frac{%
\hat{\partial}^{2}}{M}+\eta _{A}\right) \partial ^{\bar{\mu}}C-2\int B\frac{%
\hat{\partial}^{2}}{M}\left( \varsigma _{C}\frac{\hat{\partial}^{2}}{M}+\eta
_{C}\right) C
\end{equation*}%
collects evanescent contributions that are independent of the fermions.
Since the ghosts decouple, the average $\langle a\rangle $ does not give
nonevanescent one-loop contributions to $\mathcal{A}$, so we can concentrate
on the rest.

Switching to momentum space, the one-loop anomaly then reads 
\begin{equation*}
\mathcal{A}^{(1)}=\langle (S,S)\rangle _{\text{1\hspace{0.01in}loop}}=\frac{%
2g}{M}\int \frac{\mathrm{d}^{D}p}{(2\pi )^{D}}\hat{p}^{2}C(-k)\text{tr}\left[
\langle \Psi (p)\Psi ^{T}(-p+k)\rangle \left( 
\begin{tabular}{cc}
$\varsigma _{\psi }\epsilon $ & $0$ \\ 
$0$ & $-\varsigma _{\psi }^{*}\epsilon $%
\end{tabular}
\right) \right] ,
\end{equation*}
where $\Psi =(\psi _{L},\psi _{L}^{*})$. Here and below the integrals on
momenta $k$ in $\mathcal{A}^{(1)}$ are understood. These momenta can be
taken to be strictly four dimensional.

Now we expand in powers of the gauge field. By locality, power counting and
ghost number conservation, the nonevanescent contribution of the linear term
is proportional to 
\begin{equation*}
\int (\bar{\partial}^{2}C)(\partial ^{\bar{\mu}}A_{\bar{\mu}})=(S_{K},\chi
^{\prime }),\qquad \chi ^{\prime }=\frac{1}{2}\int (\partial _{\bar{\mu}}A_{%
\bar{\nu}})(\partial ^{\bar{\mu}}A^{\bar{\nu}}),
\end{equation*}%
therefore it is trivial. For the moment, we neglect the trivial
contributions and focus on the terms that are quadratic in the gauge field.
Calculating the trace and rotating the integrals to Euclidean space we find 
\begin{equation*}
\mathcal{A}^{(1)}=-8g^{3}\varepsilon _{\bar{\mu}\bar{\nu}\bar{\rho}\bar{%
\sigma}}\int C(-k_{1}-k_{2})A^{\bar{\nu}}(k_{1})A^{\bar{\sigma}%
}(k_{2})\left( k_{1}^{\bar{\mu}}k_{2}^{\bar{\rho}}\bar{I}+2k_{1}^{\bar{\mu}}%
\bar{I}^{\bar{\rho}}-2k_{2}^{\bar{\rho}}\bar{I}^{\bar{\mu}}\right) ,
\end{equation*}%
where $\varepsilon ^{0123}=1$ and $\bar{I}$, $\bar{I}^{\bar{\mu}}$ are the
finite parts of 
\begin{equation*}
I=\frac{|\varsigma _{\psi }|^{2}}{M^{2}}\int_{Eucl}\frac{\mathrm{d}^{D}p}{%
(2\pi )^{D}}\frac{(\hat{p}^{2})^{2}}{\mathcal{D}(\bar{p},k_{1},k_{2})}%
,\qquad I^{\bar{\mu}}=\frac{|\varsigma _{\psi }|^{2}}{M^{2}}\int_{Eucl}\frac{%
\mathrm{d}^{D}p}{(2\pi )^{D}}\frac{(\hat{p}^{2})^{2}\bar{p}^{\bar{\mu}}}{%
\mathcal{D}(\bar{p},k_{1},k_{2})},
\end{equation*}%
respectively, and 
\begin{equation*}
\mathcal{D}(\bar{p},k_{1},k_{2})=\tilde{D}(\bar{p})\tilde{D}(\bar{p}-k_{1})%
\tilde{D}(\bar{p}+k_{2}),\qquad \tilde{D}(\bar{q})=\bar{q}^{2}+|\varsigma
_{\psi }|^{2}\frac{(\hat{p}^{2})^{2}}{M^{2}}.
\end{equation*}%
Since $\bar{I}^{\bar{\mu}}$ is proportional to $k_{1}^{\bar{\mu}}-k_{2}^{%
\bar{\mu}}$ it is sufficient to calculate $\partial \bar{I}^{\bar{\mu}%
}/\partial k_{1}^{\bar{\mu}}$. The desired finite parts can be worked out by
inserting infrared cutoffs $\delta $ and setting $k_{1}=k_{2}=0$. The
integral over $\hat{p}$ is done first, using the standard rules of the
dimensional regularization, and factorizes an $\varepsilon =4-D$. Finally,
the integral over $\bar{p}$ is just 
\begin{equation*}
\int_{\delta }\mathrm{d}\bar{p}\hspace{0.01in}\hspace{0.01in}\bar{p}%
^{-1-\varepsilon /2}=\frac{2}{\varepsilon }+\text{finite.}
\end{equation*}%
We find 
\begin{equation*}
\bar{I}=-\frac{1}{32\pi ^{2}},\qquad \bar{I}^{\bar{\mu}}=-\frac{1}{96\pi ^{2}%
}(k_{1}^{\bar{\mu}}-k_{2}^{\bar{\mu}}),
\end{equation*}%
whence 
\begin{equation*}
\mathcal{A}^{(1)}=\frac{g^{3}}{48\pi ^{2}}\int C\varepsilon ^{\bar{\mu}\bar{%
\nu}\bar{\rho}\bar{\sigma}}F_{\bar{\mu}\bar{\nu}}F_{\bar{\rho}\bar{\sigma}%
},\qquad \langle \partial _{\mu }J^{\mu }\rangle =\frac{g^{3}}{96\pi ^{2}}%
\varepsilon ^{\bar{\mu}\bar{\nu}\bar{\rho}\bar{\sigma}}F_{\bar{\mu}\bar{\nu}%
}F_{\bar{\rho}\bar{\sigma}}-ig\left( \psi _{L}^{\ast }\frac{\delta _{l}\bar{S%
}}{\delta \psi _{L}^{\ast }}-\frac{\delta _{r}\bar{S}}{\delta \psi _{L}}\psi
_{L}\right) .
\end{equation*}%
These formulas are written at $\varepsilon =0$. Check \cite{collins} and the
appendix of ref. \cite{ABrenoYMLR} for comparison with calculations done
using the common dimensional-regularization technique.

In gauge theories with generic gauge group $G$, we insert the matrices $%
T^{a} $ and the structure constants $f^{abc}$ where appropriate, and
reintroduce the trivial contributions that we have neglected so far. In the
end, we obtain the Bardeen formula

\begin{equation}
\mathcal{A}^{(1)}=-\frac{ig^{3}}{12\pi ^{2}}\int \mathrm{d}^{D}x\
\varepsilon ^{\bar{\mu}\bar{\nu}\bar{\rho}\bar{\sigma}}\mathrm{Tr}\left[
\partial _{\bar{\mu}}C\left( A_{\bar{\nu}}\partial _{\bar{\rho}}A_{\bar{%
\sigma}}+\frac{g}{2}A_{\bar{\nu}}A_{\bar{\rho}}A_{\bar{\sigma}}\right) %
\right] +(S_{K},\chi ),  \label{a1loop}
\end{equation}
where $C=C^{a}T^{a}$, $A_{\bar{\mu}}=A_{\bar{\mu}}^{a}T^{a}$ and $\chi $ is
a local functional.

The example of this section shows that some one-loop explicit calculations
of divergent parts and anomalies with the CD regularization exhibit more or
less the same difficulties as with the ordinary dimensional technique. The
numerator algebra is considerably simplified, because the $\gamma $ matrices
are just the ones of $d$ dimensions, but denominators are not fully $%
SO(1,D-1)$ invariant. Actually, their structure is the typical one of
higher-derivative theories. However, because higher derivatives only belong
to the evanescent sector, the computational effort does not increase
dramatically.

At the same time, in section 2 we have shown that propagators are rather
involved, due to the evanescent sector, especially when gravity is present,
so in general we expect that calculations with the CD regularization are
more difficult than usual. What is important for the purposes of this paper
is that the CD\ regularization is consistent, its main virtue being that it
simplifies the proofs of all-order theorems, not that it makes explicit
calculations easier.

\section{Applications}

\setcounter{equation}{0}

In this section we outline several applications to exhibit the advantages of
the new technique. We focus on the benefits the CD regularization brings to
the proofs of all-order properties, in particular the renormalization
algorithm for general chiral gauge theories, and the Adler-Bardeen theorem.
We begin showing that it removes certain dangerous ambiguities that appear
using the common dimensional regularization. Then we prove that the CD
technique simplifies the extraction of divergent parts out of
antiparentheses of functionals, which is a key step in all renormalization
algorithms. Third, we show that it simplifies the classification of
counterterms in the presence of gravity, eliminating arbitrary dimensionless
functions that otherwise would appear in the gauge-fixing and regularization
sectors. Finally, we show that it considerably simplifies the proof of the 
\textit{manifest} Adler-Bardeen theorem for the cancellation of gauge
anomalies in perturbatively unitary gauge theories coupled to matter. By the
``manifest'' Adler-Bardeen theorem we mean the algorithm that not only
proves the cancellation of gauge anomalies to all orders, when they vanish
at one loop, but also identifies the subtraction scheme where the
cancellation occurs automatically from two loops onwards \cite{ABrenoYMLR}.
Moreover, the simplified proof obtained using the CD regularization is
suitable to be generalized to larger classes of theories, while the previous
proof is not. We also compare some features of the CD\ regularization with
Siegel's dimensional continuation \cite{siegel1,siegel2}, to further
emphasize the consistency of our technique to all orders.

We stress that, at the same time, the CD technique does preserve the good
properties of the dimensional regularization. One of them is that local
perturbative changes of field variables have Jacobian determinants
identically equal to one, which follows from the fact that the integrals of
polynomials $P(p)$ of the momenta in $\mathrm{d}^{D}p$ vanish. Thanks to
this, the Batalin-Vilkovisky master equation, as well as its deformed
version (\ref{dema}), are simpler than their general versions \cite{bata},
and several key arguments proceed more smoothly.

\subsection{Removal of ambiguities}

In the usual dimensional regularization the $\gamma $ matrices are formal
objects that satisfy the dimensionally continued Dirac algebra $\{\gamma
^{\mu },\gamma ^{\nu }\}=2\eta ^{\mu \nu }$. The completely antisymmetric
products $\gamma ^{\rho _{1}\cdots \rho _{k}}$ of $\gamma ^{\rho
_{1}},\cdots ,\gamma ^{\rho _{k}}$ are nonvanishing for arbitrary $k$, and
evanescent for $k>d$. The fermion bilinears $\bar{\psi}_{1}\gamma ^{\rho
_{1}\cdots \rho _{k}}\psi _{2}$ are all inequivalent and can be used to
build infinitely many higher-dimensional objects with the same dimensions in
units of mass, such as 
\begin{equation}
(\bar{\psi}_{1}\gamma ^{\rho _{1}\cdots \rho _{k}}\psi _{2})(\bar{\psi}%
_{3}\gamma _{\rho _{1}\cdots \rho _{k}}^{{}}\psi _{4}).  \label{4f}
\end{equation}

These properties complicate the procedure of renormalization and the proofs
of all-order theorems. For example, the Fierz identities in continued
spacetime read \cite{ABrenoYMLR} 
\begin{equation}
\psi _{2}\bar{\psi}_{3}=-\frac{1}{f(D)}\sum_{k=0}^{\infty }\frac{%
(-1)^{k(k-1)/2}}{k!}\gamma ^{\rho _{1}\cdots \rho _{k}}(\bar{\psi}_{3}\gamma
_{\rho _{1}\cdots \rho _{k}}\psi _{2}),  \label{fierz}
\end{equation}
where $f(D)=$ tr$[\mathds{1}]$, therefore we can find relations such as 
\begin{equation}
(\bar{\psi}_{1}\gamma ^{\hat{\mu}}\psi _{2})(\bar{\psi}_{3}\gamma _{\hat{\mu}%
}\psi _{4})=\frac{\varepsilon }{f(D)}(\bar{\psi}_{1}\psi _{4})(\bar{\psi}%
_{3}\psi _{2})-\frac{2}{f(D)}(\bar{\psi}_{1}\gamma ^{\hat{\rho}}\psi _{4})(%
\bar{\psi}_{3}\gamma _{\hat{\rho}}\psi _{2})-\frac{\varepsilon }{f(D)}(\bar{%
\psi}_{1}\gamma ^{\rho }\psi _{4})(\bar{\psi}_{3}\gamma _{\rho }\psi
_{2})+\cdots  \label{divam}
\end{equation}
that have the form ``fev $=$ aev $+$ fev'' and show that the distinction
between formally evanescent terms and analytically evanescent terms is
ambiguous, starting from monomials of dimension $2(d-1)$ constructed with
the fields, the sources and their derivatives.

In section 3 we have stressed that the theorem of locality of counterterms
demands that we renormalize divergent evanescences away, together with
ordinary divergences. Clearly, this statement makes sense only if we can
define divergent evanescences unambiguously. In the usual framework this is
problematic, since if we multiply, for example, both sides of formula (\ref%
{divam}) by $1/\varepsilon $ we get a relation of the type ``divev $=$
finite $+$ divev''.

In unitary, four-dimensional power counting renormalizable theories the
problem just mentioned is harmless, because it does not concern counterterms
and local contributions to anomalies \cite{ABrenoYMLR}. On the other hand,
in more general situations, such as nonrenormalizable theories, or theories
that are renormalizable by power counting, but contain higher-derivative
kinetic terms, it poses serious difficulties. A possible way out is to
define a basis of evanescent and nonevanescent monomials constructed with
the fields, the sources and their derivatives, and then express every
counterterm using that basis. This is not an easy task, since the Fierz
identities (\ref{fierz}) relate monomials that may appear to be independent
of one another at first sight.

The chiral dimensional regularization technique avoids all such troubles
from the start, because the $\gamma $ matrices and the Fierz identities are
just the usual $d$-dimensional ones, therefore the fermion bilinears are
just those we are accustomed to. In particular, they are nonevanescent and
finitely many. This is a huge simplification with respect to the ordinary
technique.

\subsection{Divergent parts of antiparentheses}

The CD regularization eludes other inconveniences, thanks to the fact that
the fields have strictly $d$-dimensional components. In particular, the
proofs of renormalizability to all orders and the manifest Adler-Bardeen
theorem \cite{ABrenoYMLR} require to extract divergent parts out of
antiparentheses like $\mathcal{A}=(\Gamma ,\Gamma )$ or $(\Gamma ,\mathcal{A}%
)$. It is helpful to know whether we can freely take this operation across
the sign of the antiparenthesis or not. We must establish, for example, that
the divergent part of $(S,\Gamma ^{(1)})$ is equal to $(S,\Gamma _{\text{div}%
}^{(1)})$, where $\Gamma ^{(1)}$ it the one-loop contribution to $\Gamma $
and $\Gamma _{\text{div}}^{(1)}$ is the divergent part of $\Gamma ^{(1)}$.
To achieve this and similar goals, we must be sure that the antiparentheses
themselves do not generate poles in $\varepsilon $, or factors of $%
\varepsilon $, and do not convert formal evanescences into analytic ones.

It is easy to prove, in full generality, that the antiparentheses do not
generate poles in $\varepsilon $. We do not repeat the derivation here,
because it is identical to the one of ref. \cite{ABrenoYMLR}. On the other
hand, using the common dimensional regularization the antiparentheses 
\textit{can} generate factors of $\varepsilon $ and convert formal
evanescences into analytic ones. Using the CD regularization this can never
happen.

More precisely, the formally evanescent quantities that appear using the CD
regularization are just $\eta ^{\hat{\mu}\hat{\nu}}$ and the evanescent
components of momenta and coordinates. The only way these objects have to
generate factors of $\varepsilon $ is by means of the contraction $\eta ^{%
\hat{\mu}\hat{\nu}}\eta _{\hat{\mu}\hat{\nu}}=-\varepsilon $. However, the
contractions of Lorentz indices brought by the functional derivatives $%
\delta /\delta \Phi ^{\alpha }$ and $\delta /\delta K_{\alpha }$ due to the
antiparentheses can never generate $\eta ^{\hat{\mu}\hat{\nu}}\eta _{\hat{\mu%
}\hat{\nu}}$, because fields and sources have no evanescent components. For
the same reason, the antiparentheses cannot convert formal evanescences into
analytic ones. Then, we can freely cross the sign of antiparentheses when we
extract divergent parts. We can write down useful symbolic identities that
summarize these properties, such as $($fev$,$fev$)=\mathrm{fev}$, $($fev$,$%
nonev$)=\mathrm{fev}$ and $($fev$,$div$)=\mathrm{divev}$, where ``nonev''
denotes convergent nonevanescent quantities, and ``div'' denotes poles in $%
\varepsilon $.

Instead, using the common dimensional regularization the metric tensor $%
g_{\mu \nu }$ and its source $K^{\mu \nu }$, for example, have traceful
evanescent components $g_{\hat{\mu}\hat{\nu}}$ and $K^{\hat{\mu}\hat{\nu}}$,
therefore the antiparentheses can generate factors of $\varepsilon $ and
convert formal evanescences into analytic ones. Moreover, four-fermion terms
can generate both factors of $\varepsilon $ and the ambiguities mentioned
above. These problems are harmless only in unitary power counting
renormalizable theories, where gravity is absent and four-fermion terms do
not appear in counterterms and local contributions to anomalies \cite%
{ABrenoYMLR}. However, in more general theories the CD\ regularization is
definitely more convenient than the usual dimensional regularization.

\subsection{Classification of counterterms in the presence of gravity}

The third application we mention concerns the classification of counterterms
and contributions to anomalies in chiral theories coupled to gravity. In
several situations, using the ordinary dimensional regularization, we may be
forced to introduce an independent metric $h_{\mu \nu }$ besides the metric
tensor $g_{\mu \nu }$ and the background metric $\bar{g}_{\mu \nu }$ around
which we expand $g_{\mu \nu }$ perturbatively. Field translations leave the
functional integral invariant, therefore correlation functions are
independent of $\bar{g}_{\mu \nu }$. Instead, they may depend on some $%
h_{\mu \nu }$, which we therefore call a ``second metric''. A second metric
can enter the classical action through the gauge fixing or the
regularization itself. Most gauge-fixing functions commonly used (e.g. $\eta
^{\rho \nu }\partial _{\rho }g_{\mu \nu }$) do introduce a second metric,
which is typically the flat-space metric $\eta _{\mu \nu }$.

When two independent metrics $g_{\mu \nu }$ and $h_{\mu \nu }$ are present,
arbitrary dimensionless functions can be built, for example functions of $%
g_{\mu \nu }h^{\mu \nu }$, $g_{\mu \nu }h^{\nu \rho }g_{\rho \sigma
}h^{\sigma \mu }$, and similar contractions. The classification of
counterterms and contributions to anomalies is then plagued with unnecessary
complications. The gauge-invariant sector is insensitive to this problem,
because general covariance forbids the arbitrary functions just mentioned.
In the gauge-fixing sector, as well as in the regularization sector,
instead, we can forbid these functions using invariance under rigid
diffeomorphisms, if the theory contains a unique metric.

Let us see an explicit example. We know that the na\"{\i}ve continuation of
the action (\ref{scf4}) to $D$ dimensions is not well regularized. One way
to deal with this problem, using the ordinary dimensional regularization, is
to introduce right-handed partners $\psi _{R}$ that decouple in the physical
limit \cite{ABrenoYMLR}. The regularized action reads 
\begin{equation}
\int \bar{\psi}_{L}i\slashed{D}\psi _{L}+\int \bar{\psi}_{R}i%
\slashed{\partial}\psi _{L}+\int \bar{\psi}_{L}i\slashed{\partial}\psi
_{R}+\int \bar{\psi}_{R}i\slashed{\partial}\psi _{R}.  \label{psir}
\end{equation}
The propagator of $\psi =\psi _{L}+\psi _{R}$ is $i/\slashed{p}$, which is
of course fine. Now, we must guarantee that the right-handed partners $\psi
_{R}$ decouple from the $S$ matrix. This is true in flat space, because by
formula (\ref{psir}) $\psi _{R}$ does not appear in any vertices, therefore
no one-particle irreducible diagrams with $\psi _{R}$ external legs can be
constructed. If we couple the theory to quantum gravity, we must
covariantize the first term of (\ref{psir}), but keep the last three terms
of (\ref{psir}) in flat space, otherwise the right-handed partners $\psi
_{R} $ do not decouple from the $S$ matrix. In this way, we do introduce the
flat-space metric $\eta _{\mu \nu }$ as a second independent metric, besides
the metric tensor $g_{\mu \nu }$.

Similarly, the most common gauge-fixing conditions for diffeomorphisms, such
as $\eta ^{\mu \rho }\partial _{\rho }g_{\mu \nu }=0$ introduce a second
metric, which is either the flat-space one or a background metric. That is
enough to create the problem we are concerned with.

As we know, a nonrenormalizable theory must be truncated so that the
truncated action $S_{T}$ contains at most a finite number of terms. When the
theory contains two metrics, we can construct infinitely many terms with the
same dimensions in units of mass, in the gauge-fixing and regularization
sectors. Then the truncation is unable to really reduce the arbitrariness to
a finite number of parameters, which makes the classification of
counterterms much less practical.

No such problems appear using the CD regularization, because it does not
introduce decoupling partners of chiral fermions, nor second metrics, and we
have proved that it is fully compatible with invariance under rigid
diffeomorphisms.

As an example, consider higher-derivative quantum gravity \cite{stelle} in
four dimensions, with or without chiral fermions. The theory is power
counting renormalizable and its action reads 
\begin{equation*}
S_{cG}=-\frac{1}{2\kappa ^{2}}\int \sqrt{|g|}(2\Lambda +R-aR_{\mu \nu
}R^{\mu \nu }-bR^{2})+\int e\bar{\psi}_{L}e_{a}^{\mu }\gamma ^{a}iD_{\mu
}\psi _{L}.
\end{equation*}%
If fermions are regularized according to (\ref{psir}), or the gauge fixing
is not invariant under rigid diffeomorphisms, the classification of
counterterms must deal with arbitrary functions that appear in the
gauge-fixing sector. In ref. \cite{stelle} this difficulty was dodged by
advocating a generalization of the Kluberg-Stern--Zuber conjecture \cite%
{kluberg}. Because of this, the proof that higher-derivative gravity is
indeed renormalizable was incomplete. It was first completed in ref. \cite%
{backc} with a different approach, using the background field method in the
absence of chiral matter. When chiral matter is present, it is much easier
to use the CD\ regularization and invariance under rigid diffeomorphisms,
because then it is sufficient to determine the coefficients of a few
Lagrangian terms. We see, again, that the CD regularization is more
convenient than the ordinary dimensional regularization.

\subsection{Proof of the \textit{manifest} Adler-Bardeen theorem}

In ref. \cite{ABrenoYMLR} the Adler-Bardeen theorem was reconsidered, and
proved in the most general four-dimensional, unitary, power counting
renormalizable theory, without using renormalization-group arguments. Thanks
to this, the subtraction scheme where gauge anomalies manifestly cancel to
all orders, if they vanish at one loop, was identified. To achieve that
goal, the dimensional regularization was merged with a suitable
higher-derivative gauge invariant regularization, and the combined technique
was called \textit{dimensional/higher-derivative} (DHD) regularization.

The DHD regularization inherits the difficulties of the ordinary dimensional
regularization. The ambiguities due to the continued Fierz identities, and
the generation of factors $\varepsilon $ by the antiparentheses make it
difficult to generalize the proof of \cite{ABrenoYMLR} to wider classes of
quantum field theories. To overcome this problem, we must first upgrade the
DHD regularization by replacing the usual dimensional regularization with
the CD regularization. We call the new combined technique \textit{%
chiral-dimensional/higher-derivative} (CDHD) regularization.

The CDHD regularization has two cutoffs, $\varepsilon $ for the CD\
regularization and $\Lambda $ for the higher-derivative regularization. The
cutoffs are removed with a procedure similar to the DHD limit defined in
ref. \cite{ABrenoYMLR}: ($i$) we first subtract the poles in $\varepsilon $,
which have nonevanescent or formally evanescent residues; ($ii$) then we
subtract the $\Lambda $ divergences, which from the $D$-dimensional
viewpoint are, again, either nonevanescent or formally evanescent; ($iii$)
then we take the analytic limit $\varepsilon \rightarrow 0$, followed by the
limit $\Lambda \rightarrow \infty $, without affecting formally evanescent
quantities; ($iv$) finally, we drop all formally evanescent quantities. We
do not give more details about the combination of the two techniques,
because everything else works exactly as in ref. \cite{ABrenoYMLR}.

The higher-derivative sector is arranged so that at fixed $\Lambda $ the
regularized theory is superrenormalizable, and has just a few one-loop,
matter-independent and source-independent divergences and potential
anomalies. Thus, at fixed $\Lambda $ the manifest Adler-Bardeen theorem is a
consequence of simple power counting arguments. At a second stage it is
proved that the theorem survives when the $\Lambda $-divergences are
renormalized away and $\Lambda $ is taken to infinity. The basic reason why
this happens is that the higher-derivative sector of the regularization is
manifestly gauge invariant.

Using the CD regularization instead of the ordinary dimensional one, the
complications mentioned in the previous subsections are properly dodged and
the upgraded version of the proof given in ref. \cite{ABrenoYMLR} proceeds
much more smoothly. However, the greatest advantage of the CDHD\
regularization is that the upgraded proof of the manifest Adler-Bardeen
theorem is ready to be generalized to a much larger class of theories. For
the moment we content ourselves with this remark, because for reasons of
space we have to postpone the investigation of this possibility to a future
publication.

\subsection{Comparison with Siegel's dimensional continuation}

Years ago, Siegel proposed a regularization technique for supersymmetric
theories \cite{siegel1} suggested by dimensional reduction, where the
dimension $d$ of spacetime is analytically continued to the complex value $D$%
, but all fields keep their physical components, and the $\gamma $ matrices
are the usual $d$-dimensional ones. Siegel's technique, called \textit{%
dimensional continuation}, was soon realized to be inconsistent by Siegel
himself \cite{siegel2}. Here we point out the differences between the CD
regularization and Siegel's dimensional continuation, to emphasize once
again how our approach works and why it is consistent to all orders. It
should also be said that our technique does not treat supersymmetry in any
special way.

If the $\gamma $ matrices are the usual $d$-dimensional ones, $\slashed{p}$
cannot depend on all components of the momentum $p$, therefore its
reciprocal is not a good propagator. Siegel's initial idea was that if we
imagine that $D$ is \textit{smaller} than $d$, $\slashed{p}$ actually does
not lose any contribution, because, so to speak, ``there are more $\gamma $
matrices than $p$ components''. However, dimensional regularization is based
on the analytic continuation of integrals to the complex plane, where no
ordering is well defined. Insisting that $D$ must be smaller than $d$ forces
us to work on the real axis, but then the analytic continuation itself is
not well defined. In four dimensions the inconsistency of Siegel's technique
can appear from four loops onwards \cite{siegel2}. At low orders
computations are safe. There, the dimensional continuation is commonly used
together with superfields \cite{grisaru}.

It is convenient to recapitulate here how the CD regularization avoids these
problems, and solves new ones that appear along the way. First, $%
D=d-\varepsilon $ and $\varepsilon $ are complex numbers, as they should be,
so $1/\slashed{p}$ is not a good propagator. To overcome this problem, we
have corrected the action by adding suitable evanescent kinetic terms.
Chiral fermions force those corrections to be higher derivative, which
brings a parameter of negative dimension (that is to say $1/M$) into the
game, even when the theory is power counting renormalizable. This is
dangerous, because an evanescent nonrenormalizable term is sufficient to
turn the theory into a nonrenormalizable one, unless there is some mechanism
that prevents this from happening.

For example, take the $\varphi ^{4}$-theory in four dimensions, and add an
evanescent $\varphi ^{6}$-vertex. The action is 
\begin{equation}
S=\int \left[ \frac{1}{2}(\partial _{\mu }\varphi )(\partial ^{\mu }\varphi
)-\frac{m^{2}}{2}\varphi ^{2}-\frac{\lambda }{4!}\varphi ^{4}-\frac{%
\varepsilon \lambda ^{\prime }}{6!M^{2}}\varphi ^{6}\right] .  \label{s6}
\end{equation}
Next, consider the one-loop diagram $G$ made with one $\varphi ^{4}$-vertex
and one $\varphi ^{6}$-vertex, and having six external legs. The factor $%
\varepsilon $ contained in the $\varphi ^{6}$-vertex simplifies the pole $%
1/\varepsilon $ of the diagram, and the result is equal to a nonevanescent
constant. The diagram $G$ is thus equivalent to a local nonevanescent
``one-loop'' vertex with six legs. Using it as a subdiagram, we can easily
construct divergent two-loop diagrams with six external legs. To renormalize
those we must make the coefficient of $\varphi ^{6}$ in (\ref{s6})
nonevanescent. In the end, $M$ propagates into the physical sector and the
theory becomes nonrenormalizable.

This cannot happen using the CD\ regularization, because weighted power
counting forbids it. For instance, the four-dimensional regularized theory 
\begin{equation}
S=\frac{1}{2}\int \left[ (\partial _{\bar{\mu}}\varphi )(\partial ^{\bar{\mu}%
}\varphi )-\frac{\varsigma _{\varphi }}{M^{2}}(\hat{\partial}^{2}\varphi
)^{2}+\frac{\eta _{\varphi }}{M}(\partial _{\hat{\mu}}\varphi )(\partial ^{%
\hat{\mu}}\varphi )-m^{2}\varphi ^{2}-\frac{\lambda }{12}\varphi ^{4}\right]
\label{sy}
\end{equation}
cannot generate nonevanescent counterterms of dimensions greater than four.
Indeed, those counterterms would have weights greater than four, so they are
excluded by weighted power counting.

By the arguments of section 3, formula (\ref{cunto2}), weighted power
counting forbids the propagation of $M$ into the physical sector also in
nonrenormalizable theories. For these reasons, despite a few coincidental
similarities, the regularization technique formulated in this paper is
rather different from Siegel's one. It does not suffer from the weaknesses
of that technique, and ultimately is fully consistent.

\section{Conclusions}

\setcounter{equation}{0}

We have formulated a modified dimensional-regularization technique that
avoids several inconveniences of the usual dimensional technique. Fields
have exactly the same components as in the physical limit, and the $\gamma $
matrices are just the usual ones. Only coordinates and momenta are continued
to complex $D$ dimensions. Invariance under rigid diffeomorphisms can be
manifestly preserved, and no extra fields are introduced for regularization
purposes, in particular no second metrics and no partners of chiral fermions
that decouple in the physical limit. We have called the new technique chiral
dimensional regularization.

The propagators are cured by means of evanescent kinetic terms. In the case
of chiral fermions such terms are higher derivative and of the Majorana
type. Weighted power counting gives us control over the locality of
counterterms and renormalization in the presence of such corrections. The
evanescent sectors of all fields must agree with one another, to produce
propagators with the same types of denominators.

Thanks to these features, the CD regularization is consistent to all orders,
and has numerous advantages. Ambiguities due to dimensionally continued
Fierz identities never show up. Divergences and evanescences propagate in a
straightforward way through the Batalin-Vilkovisky antiparentheses,
therefore the classification of counterterms and local contributions to
anomalies is considerably simpler. Moreover, the CD\ regularization is
compatible with invariance under rigid diffeomorphisms. Thanks to this
property, when gravity is present counterterms are severely constrained even
in the gauge-fixing and regularization sectors, and every truncation of the
theory contains finitely many free parameters. Finally, the proofs of
all-order theorems are less involved and ready to be generalized to wider
classes of models.

\end{document}